\documentclass[DIV=calc,paper=a4,fontsize=11pt,twocolumn]{scrartcl} 

\usepackage[english]{babel}
\usepackage[protrusion=true,expansion=true]{microtype}
\usepackage{amsmath,amsfonts,amsthm}
\usepackage{amssymb,stmaryrd,relsize}
\usepackage[final]{graphicx}
\usepackage{xcolor}
\usepackage[normal,small,hypcap,up,labelfont=bf,textfont=it]{caption}
\usepackage{epstopdf}
\usepackage{subfig}
\usepackage{booktabs}
\usepackage{fix-cm}
\usepackage{amssymb,amsfonts}
\usepackage{dsfont}
\usepackage{bbm}
\usepackage{pstricks}
\usepackage{cite}
\usepackage[utf8]{inputenc}
\usepackage[perpage,symbol*]{footmisc}
\usepackage[varg]{txfonts}
\usepackage{balance}
\usepackage{fancyhdr}
\PassOptionsToPackage{hyphens}{url}\usepackage[pdfencoding=auto,psdextra]{hyperref}
\usepackage{bookmark}
\usepackage{verbatim}
\usepackage{fontenc}
\usepackage{cuted}
\usepackage{widetext}

\usepackage{bm}
\usepackage{mathrsfs}

\DeclareCaptionFont{mycolor}{\color[HTML]{000000}}
\theoremstyle{definition}

\usepackage{accents}

\newcommand{\be}{\begin{equation}}  
\newcommand{\ee}{\end{equation}}
\newcommand{\ba}{\begin{array}}
\newcommand{\ea}{\end{array}}
\newcommand{\bea}{\begin{eqnarray}}
\newcommand{\eea}{\end{eqnarray}}

\newcommand{\bra}{\langle}
\newcommand{\ket}{\rangle}
\newcommand{\nn}{\nonumber}
\newcommand{\dtilde}[1]{\accentset{\approx}{#1}}

\usepackage{sectsty}													
\allsectionsfont{
\color[HTML]{31ADF3}\usefont{OT1}{phv}{b}{n}
}

\sectionfont{
\color[HTML]{31ADF3}\usefont{OT1}{phv}{b}{n}
}

\usepackage{fancyhdr}												
\usepackage{lettrine}
\newcommand{\initial}[1]{%
\lettrine[lines=3,lhang=0.3,nindent=0em]{
\color[HTML]{31ADF3}
{\textsf{#1}}}{}}

\usepackage{titling}															

\newcommand{\HorRule}{\color[HTML]{31ADF3}
\rule{\linewidth}{1pt}%
}

\pretitle{\vspace{-30pt} \begin{flushleft} \HorRule
\fontsize{34}{34} \usefont{OT1}{phv}{b}{n} \color[HTML]{31ADF3} \selectfont
}
\title{Steady State in Ultrastrong Coupling Regime: Expansion and First Orders}					
\posttitle{\par\end{flushleft}\vskip 0.5em}

\preauthor{\begin{flushleft}\large \lineskip 0.5em \usefont{OT1}{phv}{b}{sl} \color[HTML]{31ADF3}}
\author{Camille Lombard Latune\\[8pt]}											
\postauthor{\footnotesize \usefont{OT1}{phv}{m}{sl} \color[HTML]{000000}
Laboratoire de Physique, \'{E}cole Normale Sup\'{e}rieure de Lyon, France. E-mail: \href{mailto:cami-address@hotmail.com}{cami-address@hotmail.com}\\
Quantum Research Group, University of KwaZulu-Natal, Durban, South Africa\\
National Institute for Theoretical Physics (NITheP), Durban, South Africa\\[10pt]		
\scriptsize\usefont{OT1}{phv}{m}{n} \color[HTML]{31ADF3}{\textbf{Editors: \emph{Vinayak Jagadish}, \emph{Anton Trushechkin}, \emph{James Cresser} \& \emph{Danko D. Georgiev}} }\\[5pt]
\color[HTML]{000000}{Article history: Submitted on September 29, 2021;  Accepted on October 28, 2022; Published on November 20, 2022.}
\par\end{flushleft}\HorRule}

\date{}																				

\begin{document}
\maketitle
\thispagestyle{fancy} 			
\initial{U}\textbf{nderstanding better the dynamics and steady states of systems strongly coupled to thermal baths is a great theoretical challenge with promising applications in several fields of quantum technologies. Among several strategies to gain access to the steady state, one consists in obtaining approximate expressions of the mean force Gibbs state, the reduced state of the global system-bath thermal state, largely credited to be the steady state. Here, we present analytical expressions of corrective terms to the ultrastrong coupling limit of the mean force Gibbs state, which has been recently derived. We find that the first order term precisely coincides with the first order correction obtained from a dynamical approach---master equation in the strong-decoherence regime. This strengthens the identification of the reduced steady state with the mean force Gibbs state. Additionally, we also compare our expressions with another recent result obtained from a high temperature expansion of the mean force Gibbs state. We observe numerically a good agreement for ultra strong coupling as well as for high temperatures. This confirms the validity of all these results. In particular, we show that, in term of coherences, all three results allow one to sketch the transition from ultrastrong coupling to weak coupling.\\ Quanta 2022; 11: 53--71.}

\begin{figure}[b!]
\rule{245 pt}{0.5 pt}\\[3pt]
\raisebox{-0.2\height}{\includegraphics[width=5mm]{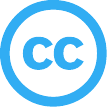}}\raisebox{-0.2\height}{\includegraphics[width=5mm]{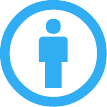}}
\footnotesize{This is an open access article distributed under the terms of the Creative Commons Attribution License \href{http://creativecommons.org/licenses/by/3.0/}{CC-BY-3.0}, which permits unrestricted use, distribution, and reproduction in any medium, provided the original author and source are credited.}
\end{figure}

\section{Introduction}  

The dynamics of quantum systems strongly coupled to thermal baths have recently received a lot of interest fuelled by hopes of understanding or even discovering new phenomena in quantum transport \cite{Chin_2012, Ribeiro_2015,Strasberg_2018,Correa_2019,Moreira_2020,Harush_2020,Dwiputra_2021, Anto_2021}, quantum thermodynamics \cite{Gelbwaser_2015,Strasberg_2016,Newman_2017,Perarnau_2018,Wertinek_2018,Newman_2020,Wiedmann_2020}, quantum sensing \cite{Correa_2017, Mehboudi_2019,Mejia_2021}, as well as understanding better the underlying physics of some essential biological functions \cite{Kolli_2012, Lambert_2013,Scholes_2017,Lambert_2020}. 
In most of these applications, to know the steady state of the system strongly coupled to the bath is often essential and sufficient. Such steady states greatly depart from usual equilibrium steady states \cite{Smith_2014,Purkayastha_2020,Cresser_2021,CLL_2021}. 
While the properties of the most general open quantum evolutions have been known for long time thanks to the seminal paper by Sudarshan, Mathews and Rau\cite{Sudarshan_1961}, their precise time evolutions and steady states are still a challenge of the theory of open quantum system \cite{FrancescoBook}.

To obtain some information about strongly coupled steady states, several strategies have been developed, including embedding techniques like reaction coordinate \cite{Garg_1985,Smith_2014,Smith_2016,Strasberg_2016} and pseudo-mode \cite{Garraway_1997,Pleasance_2017, Teretenkov_2019,Pleasance_2020}, or numerical techniques (Hierarchical Equation of Motion) \cite{Tanimura_2019,Lambert_2019,Lambert_2020}. One alternative strategy consists in focusing directly on the steady state without going through the description of the whole dynamics. In this perspective, the \emph{global} steady state of the system and bath is expected to be the global system-bath thermal state at the bath temperature \cite{Bach_2000,Frohlich_2004,Merkli_2007,Mori_2008,Konenberg_2016,Merkli_2020,Trushechkin_2022}. The steady state of the system is then given by tracing out the bath, which is often referred to as the mean force Gibbs state \cite{Cresser_2021,Trushechkin_2021, Trushechkin_2022}. Such partial trace is usually very challenging, but can be done at least approximately assuming for instance a weak coupling \cite{Mori_2008, Subasi_2012,Cresser_2021, Purkayastha_2020}. 
Another interesting regime, and potentially containing more novelty, is the ultrastrong coupling regime, when the strength of the coupling is larger than the system's energy scale. However, only few papers considered such situations. In \cite{Cresser_2021}, Cresser and Anders provide the explicit expression of the mean force Gibbs state in the limit of infinite coupling. In particular, they show that it is close to the form of the steady state obtained in \cite{Orman_2019, Goyal_2020} using arguments from eisenselection \cite{Zurek_2003}, although there are also some slight differences. Despite being an interesting result, it would be welcome to have also information on how the transition from the weak coupling limit to the ultrastrong coupling limit happens, as well as on the steady state in intermediate regimes which are experimentally more accessible.

In this perspective, introducing a technique inspired from the displaced oscillator picture \cite{Irish_2005, Li_2021} for diagonalization of the quantum Rabi model as well as reminiscent of the polaron transformation \cite{Lee_2012, Kolli_2011}, we recover the infinite coupling limit of \cite{Cresser_2021} and go beyond by providing first order corrections. In the regime of high bath temperature or low bath frequency, we derive a very simple approximate expression.  
We compare our results to two very recent derivations. The first one \cite{Trushechkin_2021} was obtained from a master equation in the so-called strong-decoherence regime, a generalization of the ultra-strong coupling regime, and actually coincides with the first order expansion derived here. This confirms that a system interacting with a thermal bath does converge, at least up to first order, to the mean force Gibbs state even in the ultra-strong coupling regime.  
The second recent derivation \cite{Timofeev_2022} consists in a high temperature expansion of the mean force Gibbs state, a generalization of a derivation introduced in \cite{Gelzinis_2020}. The comparison with this result is essentially numeric, and we find an overall good agreement in the expected regime of validity. 

Additionally, we show that the provided corrective terms allow to sketch the transition from the ultrastrong coupling limit to the weak coupling limit. Finally, we provide some higher order corrections in Section ~\ref{sec:73}.

\section{Mean force Gibbs state}

We consider the following total Hamiltonian 
\be
\label{naturalH}
{\cal H}_{SB}= H_S + H_B + \lambda H_I,
\ee
where $H_B :=\sum_k \omega_k a_k^{\dag} a_k$ is the bath Hamiltonian composed of the bosonic creation and annihilation operators $a_k^{\dag}$ and $a_k$,  $\lambda$ is a dimensionless parameter representing the strength of the system-bath coupling, $H_I :=AB$ is the system-bath coupling term, and $B := \sum_k g_k (a_k +a_k^{\dag})$ is the usual coupling bath operator.
 As in the quantum Brownian motion \cite{FrancescoBook}, there is an extra term $\lambda^2Q A^2$ appearing during the derivation of the solution which corresponds to the renormalization of the system's energies due to the interaction with the bath. This is also taken into account explicitly in \cite{Cresser_2021}. The quantity $Q$, the ``re-organization energy" \cite{May_2011, Cresser_2021}, is defined as $Q:=\int_0^{\infty} d\omega J(\omega)/\omega$, where $J(\omega)$ is the bath spectral density, $J(\omega) = \sum_k g_k^2 \delta(\omega-\omega_k)$. Such extra term is often added initially, ``by hand", so that it cancels out later on during the calculation when the renormalization takes place. Following this procedure, we add the extra term $\lambda^2 Q A^2$ so that the total Hamiltonian is now 
\be
\label{renormalizedH}
H_{SB} := H_S + \lambda^2QA^2 + \lambda AB + H_B.
\ee
Note the difference of notation between the ``natural" Hamiltonian ${\cal H}_{SB}$~\eqref{naturalH} and the renormalized Hamiltonian $H_{SB}$~\eqref{renormalizedH}.
As a matter of completeness, we also mention in Section~\ref{sec:75} the derivation starting from the ``natural'' Hamiltonian \eqref{naturalH}, which leads to renormalized final energies (or pseudo-energies). This shows in particular that both derivations are equivalent, as expected, but that the renormalization has to take place at some stage, either initially or finally. 

As mentioned in the introduction, the global steady state of the system and bath is expected to be the global thermal state (assuming $[H_S,A]\ne0$) \cite{Bach_2000,Frohlich_2004,Merkli_2007,Mori_2008,Konenberg_2016,Merkli_2020, Trushechkin_2022} 
\be
\rho_{SB}^{\textrm{th}} := Z_{SB}^{-1} e^{-\beta H_{SB}},
\ee
where $Z_{SB}:= {\textrm{Tr}}_{SB}\left(e^{-\beta H_{SB}}\right)$ is the partition function. Then, the corresponding reduced steady state is 
 \be
\rho_{S}^{\textrm{ss}} := Z_{SB}^{-1} {\textrm{Tr}}_B[e^{-\beta H_{SB}}],
\ee
which is called mean force Gibbs state \cite{Cresser_2021,Trushechkin_2021, Trushechkin_2022}.
 We are going to approximate this state in the regime of ultrastrong coupling, when $\lambda$ becomes larger than the energy scale of $H_S$. 

\section{General derivation}
\label{sec:3}

We first split $H_{SB}$ in two, $H_S$ on one hand, and $\lambda^2QA^2 + \lambda AB + H_B$ on the other hand, and take it out of the exponential using the usual identities \cite{Feynman_1951}. It leads to
\bea
\label{id1}
e^{-\beta H_{SB}} = e^{-\beta(H_B + \lambda AB +\lambda^2QA^2)}e^{-{\cal T} \int_0^{\beta} du \tilde H_S(u)},
\eea
with ${\cal T}$ representing the ``$\beta$-ordering operator'', acting on inverse temperatures in the same way as the usual time ordering operator acts on exponential time integrals, and 
\be
\tilde H_S(u) := e^{u(H_B + \lambda AB +\lambda^2QA^2)} H_S e^{-u(H_B + \lambda AB +\lambda^2QA^2)}.
\ee
Now, by noticing that
$H_B + \lambda AB +\lambda^2QA^2 = \sum_k \omega_k {\cal D}_k b_k^{\dag}b_k {\cal D}_k^{\dag}$,
where ${\cal D}_k := e^{-\frac{\lambda g_k}{\omega_k}(b_k^{\dag} - b_k)A}$ is a displacement operator by an ``amount" $\lambda g_k A/\omega_k$ acting on the mode $k$, we can re-write $H_B + \lambda AB +\lambda^2QA^2$ as a ``mixture of displaced baths". To see that, we denote by $a_l$ and $|a_l\ket$ the eigenvalues (assumed to be non-degenerate for simplicity) and corresponding eigenvectors of the observable~$A$, and by ${\cal D}_{k,l}:= e^{-\frac{\lambda g_k}{\omega_k}(b_k^{\dag} - b_k)a_l}$ the displacement operator by the quantity $\lambda g_k a_l/\omega_k$ acting only on the mode~$k$. Then, we have
\bea
 H_B + \lambda AB +\lambda^2QA^2 &=& \sum_k \omega_k {\cal D}_k b_k^{\dag}b_k {\cal D}_k^{\dag} \nn\\
 &=& \sum_k \omega_k \sum_l |a_l\ket\bra a_l|{\cal D}_k b_k^{\dag}b_k {\cal D}_k^{\dag} \nn\\
 &=& \sum_k \omega_k \sum_l |a_l\ket\bra a_l|{\cal D}_{k,l} b_k^{\dag}b_k {\cal D}_{k,l}^{\dag} \nn\\
 &=&\sum_l |a_l\ket\bra a_l| H_{B,l} ,
\eea
where we defined the ``displaced bath" $H_{B,l}:= \sum_k \omega_k {\cal D}_{k,l} b_k^{\dag}b_k {\cal D}_{k,l}^{\dag} = H_B + \lambda a_l B + \lambda^2a_l^2Q$, reminiscent of the displaced oscillator picture \cite{Irish_2005, Li_2021}. From there we obtain,
\bea
e^{u(H_B + \lambda AB +\lambda^2QA^2)} &=& e^{u \sum_l |a_l\ket\bra a_l| H_{B,l}}\nn\\
&=& \sum_l |a_l\ket\bra a_l| e^{u H_{B,l}}\label{id2} ,
\eea
so that 
\bea
\tilde H_S(u) &=& \sum_{l,l'} |a_l\ket \bra a_l| e^{uH_{B,l}} H_S |a_{l'}\ket \bra a_{l'}| e^{-uH_{B,l'}}\nn\\
&=& H_S^{\textrm{pop}} + \tilde H_S^{\textrm{coh}}(u),
\eea
where $H_S^{\textrm{pop}}:= \sum_l h_l |a_l\ket\bra a_l|$ and $\tilde H_S^{\textrm{coh}}(u):=  \sum_{l\ne l'} h_{l,l'} |a_l\ket\bra a_{l'}| e^{uH_{B,l}}e^{-uH_{B,l'}}$,
defining $h_l:= \bra a_l| H_S|a_l\ket$, and $h_{l,l'}:=\bra a_l| H_S|a_{l'}\ket$.

Taking out $H_S^{\textrm{pop}}$ from the exponential $e^{-{\cal T}\int_0^{\beta} du \tilde H_S(u)}$, we obtain,
\bea\label{id3}
e^{-{\cal T}\int_0^{\beta} du \tilde H_S(u)} &=& e^{-\beta H_S^{\textrm{pop}}} e^{-{\cal T}\int_0^\beta du \dtilde H_S^{\textrm{coh}}(u)},
\eea
with 
\bea
 \dtilde H_S^{\textrm{coh}}(u) &:=& e^{u H_S^{\textrm{pop}}} \tilde H_S^{\textrm{coh}}(u) e^{-uH_S^{\textrm{pop}}}\nn\\
 &=&  \sum_{l\ne l'} h_{l,l'} e^{u \omega_{l,l'}} |a_l\ket\bra a_{l'}|  e^{uH_{B,l}}e^{-uH_{B,l'}} ,
 \eea
 defining $\omega_{l,l'}:= h_l-h_{l'}$. Combining the identities \eqref{id1}, \eqref{id2} and \eqref{id3}, we arrive at
 \bea
 \rho_{SB}^{\textrm{th}} &=& Z_{SB}^{-1} \sum_l |a_l\ket\bra a_l| e^{-\beta H_{B,l}} e^{-\beta H_S^{\textrm{pop}}} e^{-{\cal T}\int_0^\beta du \dtilde H_S^{\textrm{coh}}(u)}\nn\\
 &=& Z_{SB}^{-1} \sum_l e^{-\beta h_l}|a_l\ket\bra a_l| e^{-\beta H_{B,l}}  e^{-{\cal T}\int_0^\beta du \dtilde H_S^{\textrm{coh}}(u)}.\nn\\
 \eea
 This leads to the following expression for the mean force Gibbs state, 
 \bea
\label{firstgen}
 \rho_S^{\textrm{ss}}  &=&Z_{SB}^{-1} \sum_l e^{-\beta h_l}|a_l\ket\bra a_l| {\textrm{Tr}}_B\left[e^{-\beta H_{B,l}}  e^{-{\cal T}\int_0^\beta du \dtilde H_S^{\textrm{coh}}(u)}\right].\nn\\
 \eea 
The toughest part is, as expected, 
\begin{widetext}
\be
\label{genexpansion}
{\textrm{Tr}}_B\left[e^{-\beta H_{B,l}}  e^{-{\cal T}\int_0^\beta du \dtilde H_S^{\textrm{coh}}(u)}\right] =  \sum_{n=0}^\infty (-1)^n\int_0^\beta du_1\int_0^{u_1}du_2...\int_0^{u_{n-1}}du_n  {\textrm{Tr}}_B\left[e^{-\beta H_{B,l}} \dtilde H_S^{\textrm{coh}}(u_1)\dtilde H_S^{\textrm{coh}}(u_2)...\dtilde H_S^{\textrm{coh}}(u_n)\right].
\ee
\end{widetext}
In the following, when we mention ``first order term'' or ``higher order terms'', we refer to the terms appearing in the above expansion \eqref{firstgen}.
So far, we made no approximation, making \eqref{firstgen} an exact expression. Since this problem is not exactly solvable, we have to make some approximations in order to reach an explicit form. Before that, let us see how one can recover the infinite coupling limit.
 
\subsection{Recovering the infinite coupling limit}
\label{sec:31}

From expression \eqref{firstgen}, it might not be obvious how one recovers the infinite coupling limit. It actually comes from the terms $\dtilde H_S^{\textrm{coh}}(u) $  which contain overlaps between displaced baths,  $e^{uH_{B,l}}e^{-uH_{B,l'}}$. The coupling strength between $S$ and each mode of the bath is given by $\lambda g_k$. Thus, when $\lambda$ goes to infinity, the displaced baths $H_{B,l}$ and $H_{B,l'}$ tend to be displaced infinitely far apart from each other (for~$l\ne l'$). Consequently, the overlap and the expectation value ${\textrm{Tr}}_B[e^{-\beta H_{B,l}}e^{uH_{B,l'}}e^{-uH_{B,l''}}]$ tends to zero for increasing coupling strength. 
 
 Extending this reasoning to higher order terms, we see that all terms ${\textrm{Tr}}_B\left[e^{-\beta H_{B,l}} \dtilde H_S^{\textrm{coh}}(u_1)\dtilde H_S^{\textrm{coh}}(u_2)...\dtilde H_S^{\textrm{coh}}(u_n)\right]$ contain multiple overlaps of different displaced baths, and therefore tends to zero as the coupling strength increases. In Section~\ref{sec:73}, we show numerically (for a two-level systems) that the higher order terms tend to zero as the coupling strength increases.
 Thus, in the infinite coupling limit, only the first term of the sum in \eqref{firstgen} is different from zero. It leads to 
  \bea
 \rho_S^{\textrm{ss}, \infty}  &=&Z_{SB}^{-1} \sum_l e^{-\beta h_l}|a_l\ket\bra a_l| {\textrm{Tr}}_B\left[e^{-\beta H_{B,l}}\right]\nn\\
  &=&(Z_S^{\textrm{ss}})^{-1}\sum_l e^{-\beta h_l}|a_l\ket\bra a_l|,\label{infcoupling}
 \eea 
 where $Z_S^{\textrm{ss}} := \frac{Z_{SB}}{Z_B}$ and $ {\textrm{Tr}}_B\left[e^{-\beta H_{B,l}}\right] =  {\textrm{Tr}}_B\left[e^{-\beta H_{B}}\right] := Z_B$ is the partition function of the uncoupled bath. Note that since $ \rho_S^{\textrm{ss}, \infty}$ is a normalized state, we also have the identity $Z_S^{\textrm{ss}} = \sum_l e^{-\beta h_l}$.
  The above expression \eqref{infcoupling} is exactly equal to the one derived in \cite{Cresser_2021}, and recently shown to coincide with the steady state in the ultrastrong coupling limit \cite{Trushechkin_2021}.
 
\subsection{First approximation}

As seen in the previous section, the term $ \dtilde H_S^{\textrm{coh}}(u) $ contains overlaps of displaced baths, so that terms ${\textrm{Tr}}_B\left[e^{-\beta H_{B,l}} \dtilde H_S^{\textrm{coh}}(u_1)\dtilde H_S^{\textrm{coh}}(u_2)...\dtilde H_S^{\textrm{coh}}(u_n)\right]$ of increasing order contain overlaps of increasing order and are therefore significantly smaller than terms of lower orders (for large coupling strength). With this assumption, we are going to retain only the first and second term,
  \bea\label{2dorderexp}
 \rho_S^{\textrm{ss}} &\simeq& Z_{SB}^{-1} \sum_l e^{-\beta h_l}|a_l\ket\bra a_l| \Bigg\{ {\textrm{Tr}}_B\left[e^{-\beta H_{B,l}}\right]\nn\\
 && - {\textrm{Tr}}_B\left[e^{-\beta H_{B,l}}  \int_0^\beta du \dtilde H_S^{\textrm{coh}}(u)\right]\Bigg\}.
 \eea 
The first term gives ${\textrm{Tr}}_B\left[e^{-\beta H_{B,l}}\right] = Z_B$, as already seen in the previous section. The second term gives,
\bea
&&{\textrm{Tr}}_B\left[e^{-\beta H_{B,l}}  \int_0^\beta du \dtilde H_S^{\textrm{coh}}(u)\right] =\sum_{l'\ne l''} h_{l',l''}  |a_{l'}\ket\bra a_{l''}|  \nn\\
&&\times\int_0^\beta du e^{u \omega_{l',l''}} {\textrm{Tr}}_B\left[e^{-\beta H_{B,l}}  e^{uH_{B,l'}}e^{-uH_{B,l''}}\right].
\eea
Since, when injected in \eqref{2dorderexp}, this expression will be multiplied by $|a_l\ket\bra a_l|$ on the left-hand side, one only needs to compute  
\bea
\sum_{l'\ne l} h_{l,l'}  |a_{l}\ket\bra a_{l'}|  \int_0^\beta du e^{u \omega_{l,l'}} {\textrm{Tr}}_B\left[e^{-\beta H_{B,l}}  e^{uH_{B,l}}e^{-uH_{B,l'}}\right],\nn\\
\eea
where the index $l$ and $l'$ were made equal. After some manipulations, we can obtain the following expression for ${\textrm{Tr}}_B\left[e^{-\beta H_{B,l}}  e^{uH_{B,l}}e^{-uH_{B,l'}}\right]$ (details provided in Section~\ref{sec:71}),
\bea
&&{\textrm{Tr}}_B\left[e^{-\beta H_{B,l}}  e^{uH_{B,l}}e^{-uH_{B,l'}}\right] \nn\\
&& = Z_B  e^{-\lambda^2a_{l',l}^2\int_0^\infty \frac{J(\omega)}{\omega^2}(e^{u\omega}-1)\left(1-\frac{1-e^{-u\omega}}{1-e^{-\omega\beta}}\right)},\label{exprTr1}
\eea
where $a_{l',l}:= a_{l'}-a_l$. We finally obtain, up to second order,
\bea
\label{final2dorder}
 \rho_S^{\textrm{ss}} &=&  \sum_l p_l^{\textrm{ss}} |a_l\ket\bra a_l|  - \sum_{l, l'; l\ne l'}  p_l^{\textrm{ss}} h_{l,l'} f_{l,l'}(\beta) |a_{l}\ket\bra a_{l'}|,\nn\\
 \eea 
with $p_l^{\textrm{ss}} := e^{-\beta h_l}/Z_S^{\textrm{ss}}$ and
\bea
\label{genf}
f_{l,l'}(\beta) 
&=& \int_0^\beta du e^{u\omega_{l,l'}} e^{-\lambda^2 a_{l',l}^2 \int_0^\infty d\omega \frac{J(\omega)}{\omega^2} (e^{u\omega} -1)\left(1-\frac{1-e^{-\omega u}}{1-e^{-\omega \beta}}\right)}.\nn\\
\eea
Note that we have the following identity (see Section~\ref{sec:72}) $p_l^{\textrm{ss}} f_{l,l'}(\beta) = p_{l'}^{\textrm{ss}}f_{l',l}(\beta)$, implying $\bra a_l|\rho_S^{\textrm{ss}}|a_{l'}\ket^* = \bra a_{l'}|\rho_S^{\textrm{ss}}|a_l\ket$, as it should be.

Additionally, $f_{+,-}(\beta)$ tends to zero when $\lambda$ goes to infinity, so that we recover the infinite coupling limit \eqref{infcoupling}. One can also see that the first order corrections affect only the coherences (in the eigenbasis of $A$). These observations coincide with the ones in \cite{Trushechkin_2021}. 
Finally, the expression obtained starting from the natural Hamiltonian \eqref{naturalH} instead of the renormalized one \eqref{renormalizedH} are the same as \eqref{final2dorder} and \eqref{genf} but substituting the pseudo-energies $h_l$ by the renormalized ones $h_l-\lambda^2a_l^2Q$ (see Section~\ref{sec:75}).

\subsection{Approximate expression of \texorpdfstring{$f_{l,l'}(\beta)$}{fl,l'(beta)}}
\label{secappf}

Depending on the bath spectral density, it might not be possible to obtain an exact analytical expression of $f_{l,l'}(\beta)$, so that some approximations would have to be made. However, assuming the bath temperature is high, or equivalently that the dominant frequencies in the bath are low, we can obtain an approximate expression of $f_{l,l'}(\beta)$ witout even specifying the form of the spectral density. 
More precisely, we assume that $J(\omega)$ vanishes for $\omega \geq \omega_c$, where $\omega_c \leq \beta^{-1}$. Then, for $\omega \in [0;\beta^{-1}]$, the factor $(e^{u\omega}-1)\left(1-\frac{1-e^{-u\omega}}{1-e^{-\omega\beta}}\right)$ can be approximated by 
\be\label{approx1}
 \frac{e^{u\omega}-1}{\omega^2}\left(1-\frac{1-e^{-u\omega}}{1-e^{-\omega\beta}}\right) = \frac{u}{\omega}\left(1-\frac{u}{\beta}\right) + {\cal O}(\omega\beta),
 \ee 
which is actually a very good approximation as soon as $\omega\beta \leq 1$.
With that we obtain
\bea\label{approx2}
 e^{-\lambda^2 a_{l',l}^2 \int_0^\infty d\omega \frac{J(\omega)}{\omega^2} (e^{u\omega} -1)\left(1-\frac{1-e^{-\omega u}}{1-e^{-\omega \beta}}\right)}  \simeq   e^{-\lambda^2a_{l',l}^2u\left(1-\frac{u}{\beta}\right)Q},\nn\\
 \eea
and
\bea\label{genapproxf}
f_{l,l'}(\beta) &\simeq&\int_0^\beta du e^{u \omega_{l,l'}}   e^{-\lambda^2a_{l',l}^2u\left(1-\frac{u}{\beta}\right)Q}\nn\\
&&\hspace{-1cm}=\frac{1}{\lambda |a_{l',l}|}\sqrt{\frac{\beta}{Q}}\Bigg\{{\textrm{DF}}\left[\frac{1}{2\lambda|a_{l',l}|}\sqrt{\frac{\beta}{Q}}(\lambda^2a_{l',l}^2Q-\omega_{l,l'})\right] \nn\\
&&+ e^{\beta\omega_{l,l'}}{\textrm{DF}}\left[\frac{1}{2\lambda|a_{l',l}|}\sqrt{\frac{\beta}{Q}}(\lambda^2a_{l',l}^2Q+\omega_{l,l'})\right]\Bigg\},\nn\\
\eea
where ${\textrm{DF}}(x):= e^{-x^2}\int_0^x {\textrm{d}}u~ e^{u^2}$ is sometimes referred to as the Dawson function. \\ 

The above expression can be further simplified as follows.
The strong coupling regime can be characterized by a re-organization energy $\lambda^2Q$ comparable to, or larger than, the energy scale of the system \cite{Cresser_2021,CLL_2021}. Thus, in the strong coupling regime one can expect to have
 $\lambda^2 Q \gg {\textrm{max}}_{l}~ |h_l|$, implying
\bea
\frac{1}{2\lambda|a_{l',l}|}\sqrt{\frac{\beta}{Q}}(\lambda^2a_{l',l}^2Q \pm \omega_{l,l'}) &\sim& \frac{1}{2\lambda|a_{l',l}|}\sqrt{\frac{\beta}{Q}}\lambda^2a_{l',l}^2Q\nn\\
& \sim& \sqrt{\beta\lambda^2Q}.
\eea
Finally, since $DF(x) = \frac{1}{2x} + {\cal O}(x^{-3})$ for $x\gg1$ (the approximation is actually very good for $x\geq 3$), we can find the following simple approximate expression for $f_{l,l'}(\beta)$ asuming $ \lambda^2 Q \gg \beta^{-1}$, 
\be\label{fsimplified}
f_{l,l'}(\beta) =\frac{1+e^{\omega_{l,l'}\beta}}{\lambda^2a_{l',l}^2Q}+\frac{\omega_{l,l'}(1-e^{\omega_{l,l'}\beta})}{\lambda^4a_{l',l}^4Q^2}+ {\cal O}[(\lambda^2Q\beta)^{-3}].
\ee

\section{Example: spin-boson model}
\label{sec:4}

As illustration of our results, we consider the versatile and famous spin-boson model \cite{Leggett_1987} characterised by the following total Hamiltonian
\be
{\cal H}_{SB} = \frac{\epsilon}{2}\sigma_z + \frac{\Delta}{2}\sigma_x + \lambda \sigma_z B + H_B,
\ee
where $\sigma_x$ and $\sigma_z$ are the Pauli matrices. The spin Hamiltonian $H_S$ can be re-written as 
\be
H_S = \frac{\epsilon}{2}\sigma_z + \frac{\Delta}{2}\sigma_x = \frac{\omega_S}{2} \big(|e\ket\bra e| -|g\ket\bra g|\big),
\ee
with $\omega_S := \sqrt{\epsilon^2+\Delta^2}$, 
\bea
&&|e\ket:= \frac{(\omega_S+\epsilon)|+\ket +\Delta |-\ket}{\sqrt{2\omega_S(\omega_S+\epsilon)}},\nn\\
&&|g\ket:= \frac{-\Delta|+\ket +(\omega_S+\epsilon) |-\ket}{\sqrt{2\omega_S(\omega_S+\epsilon)}},
\eea
and $|\pm\ket$ denotes the eigenstates of $\sigma_z$.
 Since coupling observable $A$ is equal to $\sigma_z$, we have $a_{l=\pm} = \pm 1$, $a_{l,l'} = a_{+,-} = - a_{-,+} = 2$, $ h_{l=\pm} = \pm \epsilon/2$, $h_{l,l'} = h_{+,-}=h_{-,+} = \Delta/2$, and $\omega_{l,l'} = h_l-h_{l'} = \omega_{+-}  = - \omega_{-+} = \epsilon$. Additionally, since $\sigma_z^2 = {\mathbb I}$, the renormalized Hamiltonian is equal to 
 \be
H_{SB} = \frac{\epsilon}{2}\sigma_z + \frac{\Delta}{2}\sigma_x + \lambda \sigma_z B + H_B + \lambda^2 Q,
\ee
 which simply corresponds to redefining the origin of the spin energy. This can also be verified from the renormalized pseudo-energies defined in Section~\ref{sec:75}, which are given by ${\mathbf{h}}_{l=\pm} := h_{l=\pm} - a_l^2\lambda^2Q = \pm \epsilon/2 -\lambda^2Q$. Thus, for the spin-boson model, the energy renormalization induced by the interaction with the bath has no impact on the reduced steady state. 
 
 Applying expression \eqref{final2dorder}, we obtain
 \bea\label{spinss}
 \rho_S^{\textrm{ss}} &=& \frac{1}{e^{-\beta\epsilon/2}+e^{\beta\epsilon/2}} \Big[ e^{-\beta \epsilon/2} |+\ket\bra +| + e^{\beta \epsilon/2} |-\ket\bra -| \nn\\
 &&- \frac{\Delta}{2}e^{-\beta\epsilon/2}f_{+,-}(\beta)|+\ket\bra -| - \frac{\Delta}{2}e^{\beta\epsilon/2}f_{-,+}(\beta)|-\ket\bra +|  \Big],\nn\\
 \eea
  with
  \bea
f_{+,-}(\beta) &=& \int_0^\beta du e^{\epsilon u}e^{-4\lambda^2 \int_0^\infty d\omega \frac{J(\omega)}{\omega^2} (e^{u\omega} -1)\left(1-\frac{1-e^{-\omega u}}{1-e^{\omega \beta}}\right)},\nn\\
f_{-,+}(\beta) &=& \int_0^\beta du e^{-\epsilon u}e^{-4\lambda^2 \int_0^\infty d\omega \frac{J(\omega)}{\omega^2} (e^{u\omega} -1)\left(1-\frac{1-e^{-\omega u}}{1-e^{\omega \beta}}\right)}.\nn\\
 \eea

\subsection{High temperature approximation}

Without specifying explicitly the bath spectral density, if we consider that $J(\omega)$ vanishes for frequencies smaller than $\beta^{-1}$, then the approximation \eqref{genapproxf} of Section~\ref{secappf} applies, and we have,
\bea
\label{spin1stapp}
f_{+,-}(\beta)&=&\frac{1}{2\lambda }\sqrt{\frac{\beta}{Q}}\Bigg\{{\textrm{DF}}\left[\frac{1}{4\lambda}\sqrt{\frac{\beta}{Q}}(4\lambda^2Q-\epsilon)\right] \nn\\
&&+ e^{\epsilon\beta}{\textrm{DF}}\left[\frac{1}{4\lambda}\sqrt{\frac{\beta}{Q}}(4\lambda^2Q+\epsilon)\right]\Bigg\}.
\eea
Assuming furthermore $\lambda^2Q\gg \epsilon$ and $\lambda^2Q\beta \gg1$, characteristics of the ultrastrong coupling regime, we can use expression \eqref{fsimplified}, leading to
\be\label{spin2dapp}
f_{+,-}(\beta) =\frac{1+e^{\epsilon\beta}}{4\lambda^2Q}+\frac{\epsilon(1-e^{\epsilon\beta})}{16\lambda^4Q^2}+ {\cal O}[(\lambda^2Q\beta)^{-3}].
\ee

\pagebreak
\section{Comparison with previous results}
\label{seccomparison}

In a recent paper \cite{Trushechkin_2021}, Trushechkin obtains corrections to the ultrastrong coupling limit by deriving and solving a strong-decoherence regime master equation (generalization of the ultra-strong coupling regime). Additionally, a high temperature expansion of the mean force Gibbs state was recently derived in \cite{Timofeev_2022}. 
 In this section, we compare our result with the aforementioned ones. We start by briefly introducing them.\\

\vspace{-2mm}
\noindent \emph{Trushechkin.} In \cite{Trushechkin_2021}, the author derives the first order correction to the steady state in the ultra strong coupling limit, which actually coincides with the general expressions \eqref{final2dorder} and \eqref{genf} (see details in Section~\ref{sec:76}). As a technical note, the results in \cite{Trushechkin_2021} are actually valid when the bath spectral density satisfies  $\lim_{\omega \rightarrow +\infty} \frac{J(\omega)}{\omega^2}>0$, and the equivalence of the first order correction in \cite{Trushechkin_2021} with \eqref{final2dorder} and \eqref{genf} is guaranteed only within this condition on the bath spectral density (satisfied by usual spectral densities).  

As a sanity check for the numerical simulations below, we compare, for the spin-boson model, our approximate expression \eqref{spin1stapp} of $f_{+,-}(\beta)$ with the high temperature approximation in \cite{Trushechkin_2021} based on an over-damped spectral density, also sometimes called Lorentz--Drude spectral density, 
\be
J(\omega) = \frac{2 Q}{\pi} \frac{\omega_c \omega}{\omega_c^2 + \omega^2},
\ee  
where $Q$ is precisely the re-organization energy associated with the over-damped spectral density, and $\omega_c$ represents the cutoff frequency.
Note that the dimensionless factor $\lambda$ is not explicitly present in \cite{Trushechkin_2021}, but one can make it appear by multiplying the bath spectral density by $\lambda^2$. Then, the reorganisation energy becomes $\lambda^2 Q$. 
Additionally, the steady state populations obtained in \cite{Trushechkin_2021} correspond to the one obtained for infinite coupling limit \cite{Cresser_2021}, as we also derived in \eqref{infcoupling} and \eqref{final2dorder}.\\

\vspace{-2mm}
\noindent \emph{Timofeev \& Trushechkin.}  The derivation in \cite{Timofeev_2022}, a generalization to arbitrary system-bath coupling of \cite{Gelzinis_2020}, consists in expressing the mean force Gibbs state through approximating the Hamiltonian of mean force. In other words, $\rho_S^{\textrm{ss}} = Z_{SB}^{-1}{\textrm{Tr}}_B [e^{-\beta H_{SB}}]$ is expressed in the form $Z_{MF}^{-1}e^{-\beta H_{MF}}$, and an approximated expression of $H_{MF}$, the Hamiltonian of mean force, is provided to second order in $\beta$ in the following form\cite{Timofeev_2022}
\bea
H_{MF} &=& H_S^{\textrm{diag}} - \Lambda A^2 + \sum_{n\ne m}J_{nm}e^{-\beta \Lambda (a_n-a_m)^2/6}|n\ket\bra m|\nn\\
 &&+ {\cal O}(\beta^4),
\eea
where $\Lambda$ is the re-organization energy, $A = \sum_n a_n |n\ket\bra n|$ is the system operator coupling with the bath, as previously, $H_S^{\textrm{diag}} = \sum_n \bra n| H_S |n\ket |n\ket \bra n|$, and $J_{nm} := \bra n |H_S |m\ket$. Re-expressed using the notations used throughout our paper, we have
 \bea
H_{MF} &=& \sum_l h_l|a_l\ket \bra a_l| - \lambda^2 Q A^2 \nn\\
&& + \sum_{l\ne l'}h_{l,l'}e^{-\beta \lambda^2Q (a_l-a_l')^2/6}|a_l\ket\bra a_{l'}|.
\eea
The term $- \lambda^2 Q A^2$ corresponds to the renormalization of the system's energies, which has to be performed initially as we did above, or finally, as shown in Section~\ref{sec:7}. The partition function is given by $Z_{MF} = {\textrm{Tr}}_B [e^{-\beta H_{MF}}]$.

Applying the above expression to the spin-boson model, we obtain
\bea
H_{MF} &=& \frac{\epsilon}{2}\sigma_z - \lambda^2Q + \frac{\Delta}{2}e^{-2\beta\lambda^2Q/3} \sigma_x\nn\\
&=&\frac{\epsilon}{2}\sigma_z + \frac{\Delta}{2}e^{-2\beta\lambda^2Q/3} \sigma_x,
\eea
where we dropped the renormalization term in the second line since in this situation it only corresponds to redefining the origin of the energies. Thus, the expression of the mean force Gibbs state derived in \cite{Timofeev_2022} and applied to the present spin-boson model is 
\bea
\rho_{MF} = Z_{MF}^{-1}\left[e^{-\omega_S'\beta/2}|e'\ket\bra e'| + e^{\omega_S'\beta/2}|g'\ket\bra g'|\right],
\eea
with $\omega_S' := \sqrt{\epsilon^2 + \Delta'^2}$, $\Delta' := e^{-2\beta \lambda^2Q/3}\Delta$, $|e'\ket:= \frac{(\omega_S'+\epsilon)|+\ket +\Delta' |-\ket}{\sqrt{2\omega_S'(\omega_S'+\epsilon)}}$,  $|g'\ket:= \frac{-\Delta'|+\ket +(\omega_S'+\epsilon) |-\ket}{\sqrt{2\omega_S'(\omega_S'+\epsilon)}}$, and $ Z_{MF} = e^{-\omega_S'\beta/2} + e^{\omega_S'\beta/2}$.
This expression has the merit of tending to the right limit when $\lambda \rightarrow 0$, namely $\rho_{MF} \rightarrow Z^{-1}e^{-\beta H_S}$, the usual thermal equilibrium state. This is not the case of the expression \eqref{spinss}, which is expected since it is meant to be valid in the opposite regime, when $\lambda^2Q\beta \gg1$.
%
In the eigenbasis of $\sigma_z$, the expression of $\rho_{MF}$ becomes
\bea
\label{rhoMF}
\rho_{MF} &=& \frac{1}{2}\left( 1 - \frac{\epsilon}{\omega_S'}\tanh \omega_S'\beta/2\right)|+\ket\bra +|\nn\\
&& +\frac{1}{2}\left( 1 + \frac{\epsilon}{\omega_S'}\tanh \omega_S'\beta/2\right)|-\ket\bra -| \nn\\
&& -\frac{\Delta'}{2\omega_S'}\tanh \omega_S'\beta/2 \Big(|+\ket\bra -| + |-\ket\bra +|\Big).
\eea
A brief analytical comparison between \eqref{rhoMF} and \eqref{spinss} is provided in Section~\ref{sec:74}.\\

\begin{figure}[t!]
(a)\includegraphics[width=82mm]{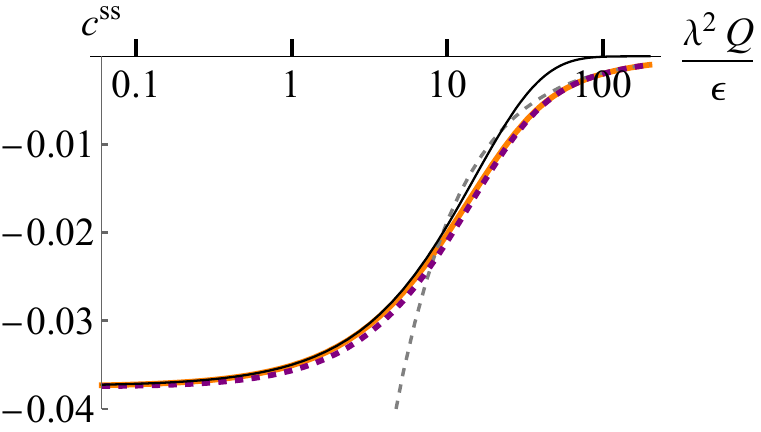}\\(b)\includegraphics[width=83.5mm]{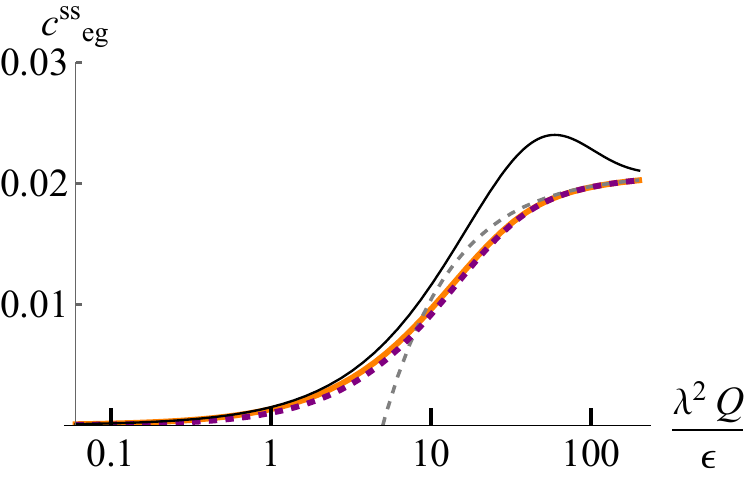}
\caption{Plots in the semi-log scale of the mean force Gibbs state coherences
(a) $c^{\textrm{ss}}:= \bra +|\rho_S^{\textrm{ss}} |-\ket$ in the eigenbasis of $A$, and (b) $c^{\textrm{ss}}_{e,g}:= \bra e|\rho_S^{\textrm{ss}} |g\ket$ in the eigenbasis of $H_S$, in function of the coupling strength $\lambda^2Q$ in unit of $\epsilon$, for $\Delta =1.5\epsilon$, $\omega_c = 2\epsilon$ and $\epsilon\beta = 0.1$. 
 The orange thick solid curve corresponds to \eqref{spinss} using the high temperature approximation \eqref{spin1stapp} of $f_{+,-}(\beta)$, while the gray thin dashed line corresponds to the further simplified expression \eqref{spin2dapp} of $f_{+,-}(\beta)$.
The purple thick dashed line is the mean force Gibbs state coherences given by the high temperature expression derived in \cite{Trushechkin_2021}. The black thin line is the mean force Gibbs state coherences given by \eqref{rhoMF}.
 }
\label{css_vs_q}
\end{figure}

\noindent In Fig.~\ref{css_vs_q}, we have plotted in semi-log scale the coherences of the mean force Gibbs state
(a) $c^{\textrm{ss}}:= \bra +|\rho_S^{\textrm{ss}} |-\ket$ in the eigenbasis of $A$, and (b) $c^{\textrm{ss}}_{e,g}:= \bra e|\rho_S^{\textrm{ss}} |g\ket$ in the eigenbasis of $H_S$, in function of the coupling strength~$\lambda^2Q$ in unit of~$\epsilon$, for $\Delta =1.5\epsilon$, $\omega_c = 2\epsilon$ and $\epsilon\beta = 0.1$. 
 The orange thick solid curve corresponds to \eqref{spinss} using the high temperature approximation \eqref{spin1stapp} of $f_{+,-}(\beta)$, while the gray thin dashed line corresponds to the further simplified expression \eqref{spin2dapp} of $f_{+,-}(\beta)$.
 The purple thick dashed line is the mean force Gibbs state coherences given by the high temperature expression derived in \cite{Trushechkin_2021}. The black thin line is the mean force Gibbs state coherences given by~\eqref{rhoMF}.

As expected, on can see an excellent agreement between \eqref{spin1stapp} (orange curve) and the expression given in \cite{Trushechkin_2021} (purple dashed line). The very slight discrepancy appearing at intermediate coupling strength comes from a slight difference in the derivation of high temperature approximation between \eqref{approx1}-\eqref{approx2} and eqs. (56)-(57) of \cite{Trushechkin_2021} (high temperature approximation before time integration of the bath correlation function).
However, there is a significant discrepancy with \eqref{rhoMF} (black line) at intermediary and strong coupling strength. 
 Beyond that, it appears from (a) that for increasing coupling strength, the mean force Gibbs state tends to a diagonal state in the eigenbasis of $A$, as expected from the ultrastrong coupling limit \eqref{infcoupling} and \cite{Orman_2019, Goyal_2020, Cresser_2021}. Conversely, for decreasing coupling strength, one can see in (b) the emergence of a progressive transition to a diagonal state in the eigenbasis of $H_S$, as expected from the weak coupling limit.

\begin{figure}[t!]
(a)\includegraphics[width=80mm]{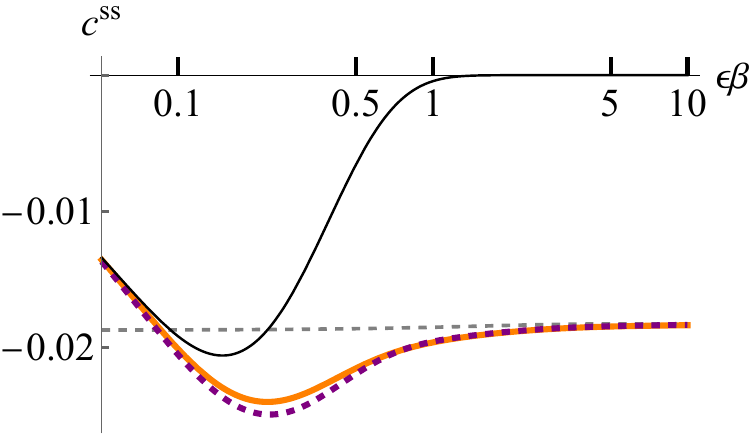}\\(b)\includegraphics[width=80mm]{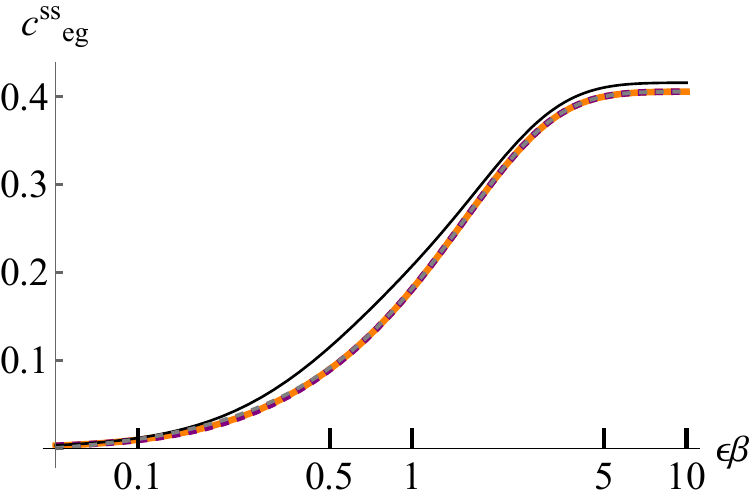}
\caption{Plots in the semi-log scale of the mean force Gibbs state coherences
(a) $c^{\textrm{ss}}:= \bra +|\rho_S^{\textrm{ss}} |-\ket$ in the eigenbasis of $A$, and (b) $c^{\textrm{ss}}_{e,g}:= \bra e|\rho_S^{\textrm{ss}} |g\ket$ in the eigenbasis of $H_S$, in function of the inverse temperature $\epsilon\beta$, for $\Delta =1.5\epsilon$, $\omega_c = 1\epsilon$ and $\lambda^2Q = 10\epsilon$. 
The orange thick solid curve corresponds to \eqref{spinss} using the high temperature approximation \eqref{spin1stapp} of $f_{+,-}(\beta)$, while the gray thin dashed line (almost indistinguishable from the orange curve) corresponds to the further simplified expression \eqref{spin2dapp} of $f_{+,-}(\beta)$.  The purple thick dashed line is the mean force Gibbs state coherences given by the high temperature expression derived in \cite{Trushechkin_2021}. The black thin line is the mean force Gibbs state coherences given by \eqref{rhoMF}. }
\label{css_vs_b}
\end{figure}

In Fig.~\ref{css_vs_b}, we have plotted in the semi-log scale of the mean force Gibbs state coherences
(a) $c^{\textrm{ss}}:= \bra +|\rho_S^{\textrm{ss}} |-\ket$ in the eigenbasis of $A$, and (b) $c^{\textrm{ss}}_{e,g}:= \bra e|\rho_S^{\textrm{ss}} |g\ket$ in the eigenbasis of $H_S$, in function of the inverse temperature $\epsilon\beta$, for $\Delta =1.5\epsilon$, $\omega_c = 1\epsilon$ and $\lambda^2Q = 10\epsilon$. 
The colour convention is the same as in the previous figure Fig.~\ref{css_vs_q}. Again, as expected, we observe a very good agreement between \eqref{spin1stapp} (orange curve) and the expression given in \cite{Trushechkin_2021} (purple dashed line). We also observe a significant discrepancy with \eqref{rhoMF} (black line) out side the high temperature regime. 

\begin{figure}[t!]
(a)\includegraphics[width=83mm]{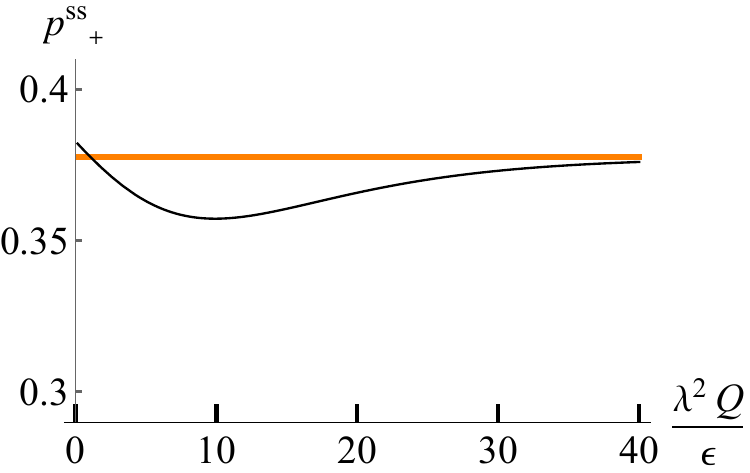}\\(b)\includegraphics[width=83mm]{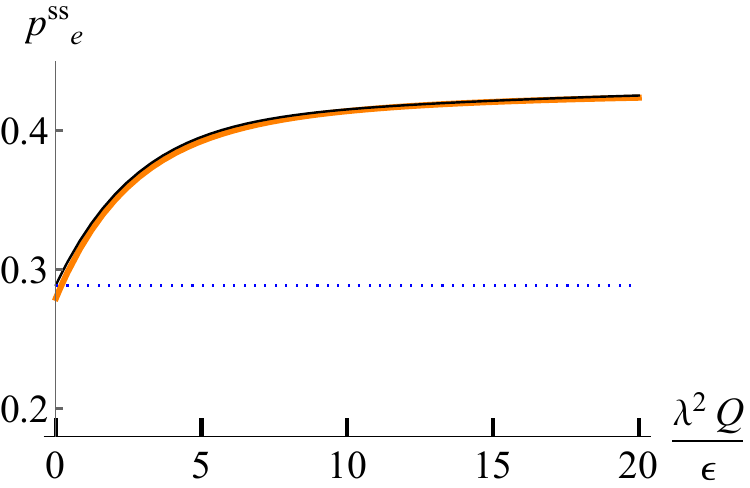}
\caption{Plots of the mean force Gibbs state excited population (a) $p_+^{\textrm{ss}}:= \bra +|\rho_S^{\textrm{ss}} |+\ket$ in the eigenbasis of $A$, and (b) $p_e^{\textrm{ss}}:= \bra e|\rho_S^{\textrm{ss}} |e\ket$ in the eigenbasis of $H_S$, in function of the coupling strength $\lambda^2Q$ in unit of $\epsilon$, for $\Delta =1.5\epsilon$ and $\epsilon\beta = 0.5$.
 The orange thick solid curve corresponds to the mean force Gibbs state excited population obtained in \eqref{spin1stapp}, \cite{Trushechkin_2021}, and \cite{Orman_2019, Goyal_2020, Cresser_2021} (all coinciding). The black thin curve corresponds to the excited population given by \eqref{rhoMF}.  The blue dotted curve corresponds to the thermal population in the vanishing coupling limit, $p_e^{\textrm{th}} = e^{-\omega_S\beta}/(1+e^{-\omega_S\beta})$. }
\label{pss_vs_q}
\end{figure}

\begin{figure}[t!]
(a)\includegraphics[width=79mm]{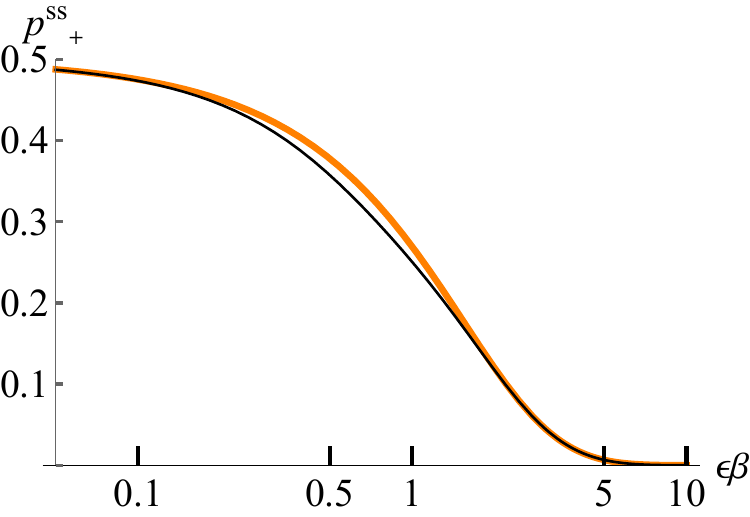}\\(b)\includegraphics[width=79mm]{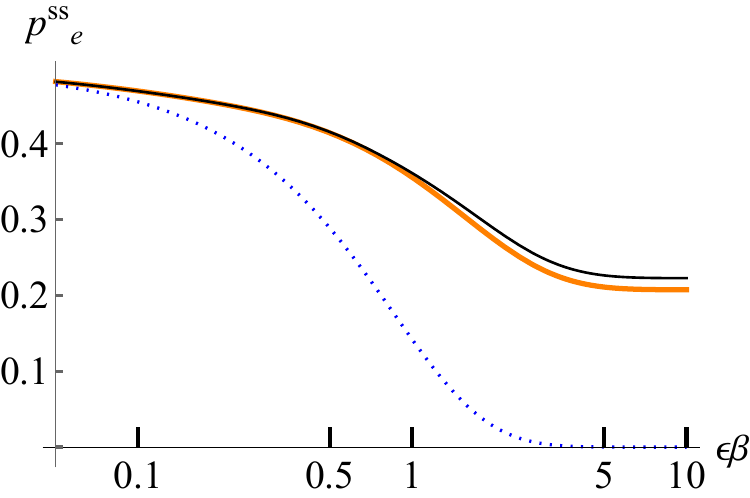}
\caption{Plots of the mean force Gibbs state excited population (a) $p_+^{\textrm{ss}}:= \bra +|\rho_S^{\textrm{ss}} |+\ket$ in the eigenbasis of $A$, and (b) $p_e^{\textrm{ss}}:= \bra e|\rho_S^{\textrm{ss}} |e\ket$ in the eigenbasis of $H_S$, in function of the inverse temperature $\epsilon\beta$ in a semi-log scale, for $\Delta =1.5\epsilon$ and $\lambda^2Q = 10\epsilon$.
The orange thick solid curve corresponds to the mean force Gibbs state excited population obtained in \eqref{spin1stapp}, \cite{Trushechkin_2021}, and \cite{Orman_2019, Goyal_2020, Cresser_2021} (all coinciding). The black thin curve corresponds to the excited population given by \eqref{rhoMF}.  The blue dotted curve corresponds to the thermal population in the vanishing coupling limit, $p_e^{\textrm{th}} = e^{-\omega_S\beta}/(1+e^{-\omega_S\beta})$.}
\label{pss_vs_b}
\end{figure}

\pagebreak
In Figs.~\ref{pss_vs_q} and \ref{pss_vs_b}, we have plotted the mean force Gibbs state excited population (a) $p_+^{\textrm{ss}}:= \bra +|\rho_S^{\textrm{ss}} |+\ket$ in the eigenbasis of $A$, and (b) $p_e^{\textrm{ss}}:= \bra e|\rho_S^{\textrm{ss}} |e\ket$ in the eigenbasis of $H_S$. 
As for the coherence, Fig.~\ref{pss_vs_q} is in function of the coupling strength $\lambda^2Q$ in unit of $\epsilon$, for $\Delta =1.5\epsilon$ and $\epsilon\beta = 0.5$, while in Fig.~\ref{pss_vs_b}, the plots are in function of the inverse temperature $\epsilon\beta$ in a semi-log scale, for $\Delta =1.5\epsilon$ and $\lambda^2Q = 5\epsilon$.
For both figures, the orange thick solid curve corresponds to the excited populations of the mean force Gibbs state obtained in \eqref{spin1stapp}, \cite{Trushechkin_2021}, and \cite{Orman_2019, Goyal_2020, Cresser_2021} (all coinciding), while  the black thin curve corresponds to the excited population given by \eqref{rhoMF}.  Finally, the blue dotted curve corresponds to the thermal population in the vanishing coupling limit, $p_e^{\textrm{th}} = e^{-\omega_S\beta}/(1+e^{-\omega_S\beta})$.

We can draw conclusions similar to Figs.~\ref{css_vs_q} and \ref{css_vs_b}, namely that both predictions coincide very well at high temperature as well as for ultra strong coupling. However, some discrepancies emerge for intermediate and large coupling strength as well as for large $\beta$.

\section{Conclusion}

We have obtained a perturbative expansion \eqref{firstgen} and \eqref{genexpansion} of the mean force Gibbs state in the ultra strong coupling regime, using a technique inspired from the displaced oscillator picture \cite{Irish_2005, Li_2021} and polaron transformation. The zero order term confirms the result of \cite{Cresser_2021, Trushechkin_2021}.
 Analytical expression of the first order term is derived, \eqref{final2dorder} and \eqref{genf}, and is found to coincide with the first order correction derived in \cite{Trushechkin_2021} (within the mild validity condition of \cite{Trushechkin_2021}, see Section~\ref{sec:76} for more detail). This increases the range of the known situations in which the steady state of a system interacting with a thermal bath does converge to the mean force Gibbs state. This convergence was recently shown for weak coupling up to the first order \cite{Mori_2008, Subasi_2012,Purkayastha_2020}, for the ultra strong coupling regime up to the zeroth order \cite{Cresser_2021}, and is now confirmed in the ultra-strong coupling regime up to the first order, thanks to the complementary results in \cite{Trushechkin_2021}. 

In the high temperature regime, a simple and explicit general expression is derived \eqref{genapproxf} for the first order corrections.  This expression is numerically compared for the spin-boson model with the result of \cite{Timofeev_2022} obtained by high temperature expansion of the mean force Gibbs state. For sanity check, a comparison of \eqref{genapproxf} with the high temperature approximation of \cite{Trushechkin_2021} is also made, and the very good agreement in all regimes of parameters is confirmed.
%
 Very good agreement is also observed at ultra strong coupling as well as high temperatures between our result and the one from \cite{Timofeev_2022}. This strengthens the validity of all three approaches. Thanks to them, we are able to draft the transition from the ultra-strong coupling regime to the weak coupling regime, Fig.~\ref{css_vs_q}b.
 
However, going further, significant discrepancies emerge between our result and \cite{Timofeev_2022} at moderate coupling strength and moderate temperatures.
For those values of parameters, further analysis with higher orders are needed in order to confirm which result is the more accurate, as well as to reconstruct accurately the full transition weak-to-ultra-strong coupling and provide a good approximation of the steady state in all regimes.  This does not seem out-of-reach according to the preliminary results on higher order terms obtained in Section~\ref{sec:73}. Additionally, benchmarking the results with other techniques like reaction coordinate or pseudo-modes would be interesting and instructive. 
 
\section{Appendix}
\label{sec:7}

\subsection{Details of the computation of \texorpdfstring{${\textrm{Tr}}_B\left[e^{-\beta H_{B,l}}  e^{uH_{B,l}}e^{-uH_{B,l'}}\right]$}{TrB[e-betaHB,l euHB,le-uHB,l']}}
\label{sec:71}

In order to compute ${\textrm{Tr}}_B\left[e^{-\beta H_{B,l}}  e^{uH_{B,l}}e^{-uH_{B,l'}}\right] $, one can first rearrange it in the following way,
\bea
&&{\textrm{Tr}}_B\left[e^{-\beta H_{B,l}}  e^{uH_{B,l}}e^{-uH_{B,l'}}\right] = {\textrm{Tr}}_B\left[e^{-(\beta-u) H_{B,l}} e^{-uH_{B,l'}}\right]\nn\\
 &&\hspace{0.5cm}= {\textrm{Tr}}_B\left[\underset{k}{\otimes} \left({\cal D}_{k,l} e^{-(\beta-u) b_k^{\dag}b_k}{\cal D}^{\dag}_{k,l}\right) e^{-uH_{B,l'}}\right]\nn\\
&&\hspace{0.5cm}= {\textrm{Tr}}_B\left[e^{-(\beta-u) H_{B}} e^{-u\sum_k \omega_k {\cal D}_{k,l}^{\dag}{\cal D}_{k,l'}b_k^{\dag}b_k{\cal D}_{k,l'}^{\dag}{\cal D}_{k,l}}\right]\nn\\
&&\hspace{0.5cm}= {\textrm{Tr}}_B\left[e^{-(\beta-u) H_{B}} e^{-u(H_B + \lambda a_{l',l} B + \lambda^2 a_{l',l}^2 Q)}\right],
\eea
where $a_{l',l}:= a_{l'}-a_l$. ``Taking out" of the second exponential the Hamiltonian $H_B$, we obtain,
\bea
\label{split1}
 &&{\textrm{Tr}}_B\left[e^{-(\beta-u) H_{B}} e^{-u(H_B + \lambda a_{l',l} B + \lambda^2 a_{l',l}^2 Q)}\right] \nn\\
 &&= e^{-u (\lambda^2 a_{l',l}^2 Q)}{\textrm{Tr}}_B\left[e^{-\beta H_{B}} e^{- \lambda a_{l',l}{\cal T}\int_0^u ds \tilde B(s) }\right],
 \eea
with $\tilde B(s) := e^{s H_B} B e^{-s H_B} = \sum_k g_k (e^{s\omega_k}b_k^{\dag} +e^{-s\omega_k}b_k)$. The time ordered operator $e^{- \lambda a_{l',l}{\cal T}\int_0^u ds \tilde B(s) }$ can be split in two as follows, 
\be\label{split2}
e^{- \lambda a_{l',l}{\cal T}\int_0^u ds \tilde B(s) } = e^{-\lambda^2a_{l',l}^2 f(u)} e^{\lambda a_{l',l} {\cal B}(-u)}e^{-\lambda a_{l',l} {\cal B}^{\dag}(u)},
\ee
with $f(u) := -u Q + \sum_k \frac{g_k^2}{\omega_k^2}(e^{u\omega_k}-1)$, ${\cal B}(-u) := \sum_k \frac{g_k}{\omega_k}(e^{-u\omega_k}-1)b_k$, and ${\cal B}^{\dag}(u) := \sum_k \frac{g_k}{\omega_k}(e^{u\omega_k}-1)b_k^{\dag}$. The above decomposition can be demonstrated as follows. We denote the left-hand side of \eqref{split2} as $L(u):=e^{- \lambda a_{l',l}{\cal T}\int_0^u ds \tilde B(s) } $ and the right-hand side by $R(u):= e^{-\lambda^2a_{l',l}^2 f(u)} e^{\lambda a_{l',l} {\cal B}(-u)}e^{-\lambda a_{l',l} {\cal B}^{\dag}(u)}$. 
By definition of the time ordering operator, or alternatively by taking the time derivative with respect to $u$, the left-hand side of \eqref{split2} satisfies the differential equation
\be
\label{eqdiffR}
\frac{\partial}{\partial u} L(u) =  -\lambda a_{l',l}\tilde B(u) L(u).
\ee
Now, by taking the time derivative with respect to $u$ of the right-hand side of \eqref{split2}, one obtains the following differential equation
\bea
\frac{\partial}{\partial u} R(u) &=& \left[-\lambda^2 a_{l',l}^2 \frac{\partial}{\partial u} f(u) + \lambda a_{l',l}\frac{\partial}{\partial u}{\cal B}(-u)\right] R(u) \nn\\
&&-\lambda a_{l',l} R(u) \frac{\partial}{\partial u}{\cal B}^{\dag}(u).
\eea
Using the expressions of $f(u)$, ${\cal B}(u)$, and ${\cal B}^{\dag}(u)$, we obtain,
\bea
\frac{\partial}{\partial u} f(u) &=& -Q +\sum_k\frac{g_k^2}{\omega_k}e^{u\omega_k}, \nn\\
\frac{\partial}{\partial u}{\cal B}(-u) &=& -\sum_k g_k e^{-u\omega_k}b_k, \nn\\
\frac{\partial}{\partial u}{\cal B}^{\dag}(u) &=& \sum_k g_k e^{u\omega_k}b_k^{\dag}.
\eea
Then, with the help of the identity 
\be
R(u) \frac{\partial}{\partial u}{\cal B}^{\dag}(u) = \left[\frac{\partial}{\partial u}{\cal B}^{\dag}(u) + \lambda a_{l',l}\sum_k \frac{g_k^2}{\omega_k}(1-e^{u\omega_k})\right]R(u),
\ee
and the discrete version of the re-organization energy $Q= \sum_k \frac{g_k^2}{\omega_k} \equiv \int_0^{\infty} d\omega \frac{J(\omega)}{\omega}$, we arrive at
\bea
\frac{\partial}{\partial u} R(u) &=& -\lambda a_{l',l}\left[ \sum_k g_k \left(e^{u\omega_k}b_k^{\dag} + e^{-u\omega_k}b_k\right)\right] R(u),\nn\\
\eea
which is exactly the same differential equation as \eqref{eqdiffR}. Since the initial conditions are the same, namely $R(0) = L(0)= \mathbb{I}$, we conclude that $R(u)=L(u)$ for all times $u$. 

\begin{widetext}
A simple way to conclude the computation of the trace \eqref{split1} is using the normal order representation of the operators. Using the following formula \cite[p.~116]{Louisell_1964}
\be
e^{\alpha a^{\dag} a} = \sum_{n=0}^\infty \frac{(e^\alpha-1)^n}{n!}(a^{\dag})^n a^n,
\ee
where $\alpha$ is a c-number and $a^{\dag}$, $a$ are bosonic operators, we obtain,
\bea
 &&{\textrm{Tr}}_B\left[e^{-(\beta-u) H_{B}} e^{-u(H_B + \lambda a_{l',l} B + \lambda^2 a_{l',l}^2 Q)}\right] \nn\\
 && \hspace{2cm}= e^{-u (\lambda^2 a_{l',l}^2 Q)}e^{-\lambda^2a_{l',l}^2 f(u)} {\textrm{Tr}}_B\left[e^{-\lambda a_{l',l} {\cal B}^{\dag}(u)}\left(\otimes_k \sum_{n=0}^\infty \frac{(e^{-\beta\omega_k}-1)^n}{n!}(b_k^{\dag})^n b_k^n \right) e^{\lambda a_{l',l} {\cal B}(-u)}\right]\label{standardform}\\
 &&\hspace{2cm}= e^{- \lambda^2 a_{l',l}^2 \sum_k \frac{g_k^2}{\omega_k^2}(e^{u\omega_k}-1)} \prod_k\left( \int_{\mathbb{C}} \frac{{\textrm{d}}^2\alpha_k}{\pi} e^{-\lambda a_{l',l} \frac{g_k}{\omega_k}(e^{u\omega_k}-1)\alpha_k^{*}} \sum_{n=0}^\infty \frac{(e^{-\beta\omega_k}-1)^n}{n!}|\alpha_k|^{2n} e^{\lambda a_{l',l} \frac{g_k}{\omega_k}(e^{-u\omega_k}-1)\alpha_k}\right)\label{tracecoh}\\
 &&\hspace{2cm}=  e^{- \lambda^2 a_{l',l}^2 \sum_k \frac{g_k^2}{\omega_k^2}(e^{u\omega_k}-1)} \prod_k\left( \int_{\mathbb{C}} \frac{{\textrm{d}}^2\alpha_k}{\pi} e^{-\lambda a_{l',l} \frac{g_k}{\omega_k}(e^{u\omega_k}-1)\alpha_k^{*}} e^{(e^{-\beta\omega_k}-1)|\alpha_k|^2} e^{\lambda a_{l',l} \frac{g_k}{\omega_k}(e^{-u\omega_k}-1)\alpha_k}\right)\nn\\
 &&\hspace{2cm}=  e^{- \lambda^2 a_{l',l}^2 \sum_k \frac{g_k^2}{\omega_k^2}(e^{u\omega_k}-1)} \prod_k\left( \int_{\mathbb{R}^2} \frac{{\textrm{d}}x_k{\textrm{d}}y_k}{\pi} e^{ -\lambda a_{l',l} \frac{g_k}{\omega_k}(e^{u\omega_k}-1)(x_k-iy_k)  +(e^{-\beta\omega_k}-1)(x_k^2+y_k^2)+ \lambda a_{l',l} \frac{g_k}{\omega_k}(e^{-u\omega_k}-1)(x_k+iy_k)}\right)\nn\\
 &&\hspace{2cm}=  e^{- \lambda^2 a_{l',l}^2 \sum_k \frac{g_k^2}{\omega_k^2}(e^{u\omega_k}-1)} \prod_k\left( \int_{\mathbb{R}^2} \frac{{\textrm{d}}x_k{\textrm{d}}y_k}{\pi} e^{(e^{-\beta\omega_k}-1)(x_k^2+y_k^2)  + \lambda a_{l',l} \frac{g_k}{\omega_k}(e^{-u\omega_k}-e^{u\omega_k})x_k + i\lambda a_{l',l} \frac{g_k}{\omega_k}(e^{-u\omega_k}+e^{u\omega_k}-2)y_k   }\right)\nn\\
  &&\hspace{2cm}=  e^{- \lambda^2 a_{l',l}^2 \sum_k \frac{g_k^2}{\omega_k^2}(e^{u\omega_k}-1)} \prod_k \left(\frac{1}{1-e^{-\omega_k\beta}}\right) e^{-\lambda^2a_{l',l}^2\sum_k \frac{g_k^2}{\omega_k^2}\frac{2-e^{u\omega_k}-e^{-u\omega_k}}{1-e^{-\omega_k\beta}}}\nn\\
  &&\hspace{2cm}= Z_B  e^{-\lambda^2a_{l',l}^2\sum_k \frac{g_k^2}{\omega_k^2}(e^{u\omega_k}-1)\left(1-\frac{1-e^{-u\omega_k}}{1-e^{-\omega_k\beta}}\right)}\nn\\
   &&\hspace{2cm}= Z_B  e^{-\lambda^2a_{l',l}^2\int_0^\infty \frac{J(\omega)}{\omega^2}(e^{u\omega}-1)\left(1-\frac{1-e^{-u\omega}}{1-e^{-\omega\beta}}\right)},\label{exprTr2}
 \eea
where $Z_B:=\Pi_k \left(\frac{1}{1-e^{-\omega_k\beta}}\right) = {\textrm{Tr}}_B\left(e^{-\beta H_B}\right)$. In line \eqref{tracecoh}, we used the property of the coherent states $|\alpha_k\ket$ which form an over-complete basis of the Hilbert space of the $k$-mode, so that the trace of any operator ${\cal O}_k$ acting in that Hilbert space can be computed as ${\textrm{Tr}}_k({\cal O}_k) = \int_{\mathbb{R}^2}\frac{{\textrm{d}}\alpha_k^2}{\pi}\bra \alpha_k|{\cal O}_k|\alpha_k\ket$. Additionally, we expressed in the last lines the complex variable $\alpha_k$ explicitly in term of its real and imaginary part $\alpha_k = x_k +iy_k$.

\subsection{Important property of \texorpdfstring{$f_{l,l'}(\beta)$}{fl,l'(beta)}}
\label{sec:72}

By applying the change of variable $v=\beta-u$, one can re-write $f_{l,l'}(\beta)$ in the form
\bea
f_{l,l'}(\beta) &=& \int_0^\beta du e^{u\omega_{l,l'}} e^{-\lambda^2 a_{l',l}^2 \int_0^\infty d\omega \frac{J(\omega)}{\omega^2} \frac{e^{\omega\beta/2}+e^{-\omega\beta/2}-e^{\omega (u-\beta/2)}-e^{-\omega(u-\beta/2)}}{e^{\omega\beta/2}-e^{-\omega \beta/2}}} \nn\\
&=& e^{\beta \omega_{l,l'}} \int_0^\beta dv e^{-v\omega_{l,l'}} e^{-\lambda^2 a_{l',l}^2 \int_0^\infty d\omega \frac{J(\omega)}{\omega^2} \frac{e^{\omega\beta/2}+e^{-\omega\beta/2}-e^{\omega (v-\beta/2)}-e^{-\omega(v-\beta/2)}}{e^{\omega\beta/2}-e^{-\omega \beta/2}}}.
\eea
Then, one can easily verify that $p_l^{\textrm{ss}} f_{l,l'}(\beta) = p_{l'}^{\textrm{ss}}f_{l',l}(\beta)$. This implies in particular that $\bra a_l|\rho_S^{\textrm{ss}}|a_{l'}\ket^* = \bra a_{l'}|\rho_S^{\textrm{ss}}|a_l\ket$, as required by the Hermicity of $\rho_S^{\textrm{ss}}$.

\subsection{Higher order terms}
\label{sec:73}

As explained in Section~\ref{sec:3}, we consider only the first two terms of the infinite sum in Eq.~\eqref{genexpansion}. In Section~\ref{sec:31}, we gave an intuitive argument to justify that the second term as well as all following terms in Eq.~\eqref{genexpansion} converge to zero when the coupling strength increases. We now come back to this point an provide a more rigorous argument. 

For simplicity, we focus here on the spin-boson model used in Section~\ref{sec:4}, while this could be easily extended to arbitrary systems. . 
Combining Eqs.~\eqref{firstgen} and \eqref{genexpansion}, we start re-writing the exact expression of $\rho_S^{\textrm{ss}}$ as 
\be
\rho_S^{\textrm{ss}} = (Z_{S}^{ss})^{-1} \left[e^{-\beta h_+}\sum_{n=0}^{\infty} T_{+,n}+e^{-\beta h_-}\sum_{n=0}^{\infty} T_{-,n}\right],
\ee
with $h_\pm = \bra \pm|H_S|\pm\ket = \pm \epsilon/2$, and where 
\be
\label{exprTln}
T_{\pm,n} := (-1)^n\int_0^\beta du_1\int_0^{u_1}du_2...\int_0^{u_{n-1}}du_n 
|\pm\ket\bra \pm|{\textrm{Tr}}_B\left[Z_B^{-1}e^{-\beta H_{B,\pm}} \dtilde H_S^{\textrm{coh}}(u_1)\dtilde H_S^{\textrm{coh}}(u_2)...\dtilde H_S^{\textrm{coh}}(u_n)\right],
\ee
 with $H_{B,\pm} := H_B \pm \lambda B + \lambda^2Q$, $T_{\pm,0} = |\pm\ket\bra \pm|$ and 
 \be
 Z_{S}^{ss}:= \frac{Z_{SB}}{Z_B} = e^{-\beta \epsilon/2}\sum_{n=0}^{\infty}\bra +|T_{+,n}|+\ket +e^{\beta \epsilon/2}\sum_{n=0}^{\infty}\bra -|T_{-,n}|-\ket.
 \ee
We then work on the integrand 
\bea
t_{\pm,n}&:=&|\pm\ket\bra \pm|{\textrm{Tr}}_B\left[e^{-\beta H_{B,\pm}} \dtilde H_S^{\textrm{coh}}(u_1)\dtilde H_S^{\textrm{coh}}(u_2)...\dtilde H_S^{\textrm{coh}}(u_n)\right].
\eea
Starting with $t_{+,n}$, we have
\bea
t_{+,n}&=& |+\ket\bra +|{\textrm{Tr}}_B\left[Z_B^{-1}e^{-\beta H_{B,+}} \dtilde H_S^{\textrm{coh}}(u_1)\dtilde H_S^{\textrm{coh}}(u_2)...\dtilde H_S^{\textrm{coh}}(u_n)\right]\nn\\
&=&Z_B^{-1} |+\ket\bra +|{\textrm{Tr}}_B\Bigg[e^{-\beta H_{B,+}} \left( h_{+,-}e^{u_1\omega_{+,-}}|+\ket\bra -|e^{u_1H_{B,+}}e^{-u_1H_{B,-}} +h_{-,+}e^{u_1\omega_{-,+}}|-\ket\bra +|e^{u_1H_{B,-}}e^{-u_1H_{B,+}} \right)\nn\\
&& \hspace{2.5cm}\times\dtilde H_S^{\textrm{coh}}(u_2)...\dtilde H_S^{\textrm{coh}}(u_n) \Bigg]\nn\\
&=&Z_B^{-1} h_{+,-}e^{u_1\omega_{+,-}}|+\ket\bra -|{\textrm{Tr}}_B\Bigg[ e^{-(\beta-u_1) H_{B,+}}e^{-u_1H_{B,-}} \dtilde H_S^{\textrm{coh}}(u_2)...\dtilde H_S^{\textrm{coh}}(u_n) \Bigg]\nn\\
&=& Z_B^{-1} h_{+,-}e^{u_1\omega_{+,-}}|+\ket\bra -|{\textrm{Tr}}_B\Bigg[e^{-(\beta-u_1) H_{B,+}}e^{-u_1H_{B,-}} \nn\\
&&\hspace{2.5cm}\times\left( h_{+,-}e^{u_2\omega_{+,-}}|+\ket\bra -|e^{u_2H_{B,+}}e^{-u_2H_{B,-}} +h_{-,+}e^{u_2\omega_{-,+}}|-\ket\bra +|e^{u_2H_{B,-}}e^{-u_2H_{B,+}} \right)\dtilde H_S^{\textrm{coh}}(u_3)...\dtilde H_S^{\textrm{coh}}(u_n) \Bigg]\nn\\
&=& Z_B^{-1} h_{+,-}h_{-,+}e^{u_1\omega_{+,-}}e^{u_2\omega_{-,+}}|+\ket\bra +|{\textrm{Tr}}_B\Bigg[e^{-(\beta-u_1) H_{B,+}}e^{-(u_1-u_2)H_{B,-}}e^{-u_2H_{B,+}} \dtilde H_S^{\textrm{coh}}(u_3)...\dtilde H_S^{\textrm{coh}}(u_n) \Bigg]\nn\\
&=& Z_B^{-1}
\left(\frac{\Delta}{2}\right)^n e^{\epsilon (u_1-u_2+ ... -(-1)^n u_n)}\nn\\
&&\hspace{2.5cm}\times{\textrm{Tr}}_B\Bigg[e^{-(\beta-u_1) H_{B,+}} e^{-(u_1-u_2)H_{B,-}}  ...    e^{-(u_{n-1}-u_n)H_{B,(-1)^{n-1}}}e^{-u_nH_{B,(-1)^n}} \Bigg]|+\ket\bra (-1)^n|,
\eea
since $h_{+,-}=h_{-,+}=\Delta/2$ and $\omega_{+,-} = -\omega_{-,+} = \epsilon$.
Reproducing the decomposition used in Eq.\eqref{split1} and \eqref{split2}, we obtain
\bea
e^{-\alpha_i H_{B,\pm}} = e^{- \lambda^2\bar f(\alpha_i)} e^{-\alpha_i H_B} e^{\pm\lambda {\cal B}(-\alpha_i)}e^{\mp\lambda {\cal B}^{\dag}(\alpha_i)} \label{splitting}
\eea
with $\bar f(\alpha_i) := \sum_k \frac{g_k^2}{\omega_k^2}(e^{\alpha_i\omega_k}-1)$ and $\alpha_i = u_i-u_{i+1}$. The aim is now to re-write 
\be
e^{-(\beta-u_1) H_{B,+}} e^{-(u_1-u_2)H_{B,-}}  ...    e^{-(u_{n-1}-u_n)H_{B,(-1)^{n-1}}}e^{-u_nH_{B,(-1)^n}}
\ee
in normal order so that we can compute the trace. To have a better understanding of how we will proceed, we first consider the simpler situation where $n=1$. We have, using \eqref{splitting} twice,
\be
e^{-(\beta-u_1) H_{B,+}} e^{-u_1H_{B,-}}
=  e^{- \lambda^2\bar f(\beta-u_1)}e^{- \lambda^2\bar f(u_1)}
e^{-(\beta-u_1) H_B} e^{\lambda {\cal B}(-\beta+u_1)}e^{-\lambda {\cal B}^{\dag}(\beta-u_1)}
e^{-u_1 H_B} e^{-\lambda {\cal B}(-u_1)}e^{\lambda {\cal B}^{\dag}(u_1)}. \label{n1}
\ee
Using the properties
\bea
&&e^{\chi_k b_k}e^{-\alpha_i \omega_k b_k^{\dag}b_k} = e^{-\alpha_i \omega_k b_k^{\dag}b_k}e^{\chi_k e^{-\alpha_i\omega_k} b_k}\nn\\
&&e^{\chi_k b_k^{\dag}}e^{-\alpha_i \omega_k b_k^{\dag}b_k} = e^{-\alpha_i \omega_k b_k^{\dag}b_k}e^{\chi_k e^{\alpha_i\omega_k} b_k^{\dag}}
\eea
for arbitrary coefficient $\chi_k$, we have
\be
e^{-(\beta-u_1) H_{B,+}} e^{-u_1H_{B,-}} = e^{- \lambda^2\bar f(\beta-u_1)}e^{- \lambda^2\bar f(u_1)}e^{-\beta H_B}
e^{\lambda \sum_k\frac{g_k}{\omega_k}e^{-u_1\omega_k}(e^{-(\beta-u_1)\omega_k}-1)b_k}
e^{-\lambda \sum_k\frac{g_k}{\omega_k}e^{u_1\omega_k}(e^{(\beta-u_1)\omega_k}-1)b_k^{\dag}}
e^{-\lambda {\cal B}(-u_1)}e^{\lambda {\cal B}^{\dag}(u_1)}.
\ee
We finally use the identity 
\be
e^{\chi_k b_k^{\dag}}e^{\xi_k b_k} =e^{-\chi_k\xi_k} e^{\xi_k b_k}e^{\chi_k b_k^{\dag}},
\ee
to obtain,
\bea
e^{-(\beta-u_1) H_{B,+}} e^{-u_1H_{B,-}} &=& e^{- \lambda^2\bar f(\beta-u_1)}e^{- \lambda^2\bar f(u_1)}
e^{-\lambda^2\sum_k \frac{g_k^2}{\omega_k^2}e^{u_1\omega_k}(e^{(\beta-u_1)\omega_k}-1)(e^{-u_1\omega_k}-1)}
e^{-\beta H_B} \nn\\
&&~~~ \times
e^{\lambda\sum_k\frac{g_k}{\omega_k}[ e^{-u_1\omega_k}(e^{-(\beta-u_1)\omega_k}-1) - (e^{-u_1\omega_k}-1) ] b_k}
e^{-\lambda\sum_k\frac{g_k}{\omega_k}[e^{u_1\omega_k}(e^{(\beta-u_1)\omega_k}-1) - (e^{u_1\omega_k}-1)]b_k^{\dag}}. \label{endn1}
\eea
The above expression is not yet in normal order, but once injected in the trace, one can use the permutation property as in the previous section to obtain an expression in normal order, exactly of the same form as in \eqref{standardform}, the only difference being the coefficients in front of the annihilation and creation operators $b_k$ and $b_k^{\dag}$.  We then proceed to the computation of the trace exactly in the same way as in the previous section. We obtain something of the form,
\be
\label{psiform}
{\textrm{Tr}}_B\Bigg[e^{-(\beta-u_1) H_{B,+}} e^{-u_1H_{B,-}} \Bigg] = Z_B  e^{-\lambda^2 \Psi_{+,-}(\beta,u_1) },
\ee
where $\Psi_{+,-}(\beta,u_1)$ is a real function.
 
If now we consider the term $n=2$, we can use the above form for $e^{-(\beta-u_1) H_{B,+}} e^{-(u_1-u_2)H_{B,-}}$, and then split $ e^{-u_2H_{B,+}}$ using \eqref{splitting}. We obtain an expression of the same form as \eqref{n1}, and thus one can repeat the above steps to reach an expression of the later form \eqref{psiform}, namely,
\be
{\textrm{Tr}}_B\Bigg[e^{-(\beta-u_1) H_{B,+}} e^{-(u_1-u_2)H_{B,-}} e^{-u_2H_{B,+}} \Bigg]
 = Z_B e^{-\lambda^2 \Psi_{+,-,+}(\beta,u_1,u_2) }.
\ee
 We can repeat the same procedure for any $n$, obtaining
\bea\label{psiformgen}
{\textrm{Tr}}_B\Bigg[e^{-(\beta-u_1) H_{B,+}} e^{-(u_1-u_2)H_{B,-}}  ...    e^{-(u_{n-1}-u_n)H_{B,(-1)^{n-1}}}e^{-u_nH_{B,(-1)^n}} \Bigg] =Z_B  e^{-\lambda^2 \Psi_{+,-,+,...,(-1)^n}(\beta,u_1,u_2,...u_n) },
\eea
where $\Psi_{+,-,+,...,(-1)^n}(\beta,u_1,u_2,...u_n)$ is a real function. 

Having this result in mind, one can obtain a more direct form of $\Psi_{+,-,+,...,(-1)^n}(\beta,u_1,u_2,...u_n)$ simply by computing
\be
\frac{\partial^2}{\partial \lambda^2} {\textrm{Tr}}_B\Bigg[e^{-(\beta-u_1) H_{B,+}} e^{-(u_1-u_2)H_{B,-}}  ...    e^{-(u_{n-1}-u_n)H_{B,(-1)^{n-1}}}e^{-u_nH_{B,(-1)^n}} \Bigg] _{|\lambda=0}.
\ee
After computing this second derivative and evaluating it in $\lambda = 0$, followed by some simple algebraic manipulations, we arrived at 
\bea
\Psi_{+,-,+,...,(-1)^n}(\beta,u_1,u_2,...u_n) &=& -\frac{1}{2}\frac{\partial^2}{\partial \lambda^2} {\textrm{Tr}}_B\Bigg[ Z_B^{-1}e^{-(\beta-u_1) H_{B,+}} e^{-(u_1-u_2)H_{B,-}}  ...    e^{-(u_{n-1}-u_n)H_{B,(-1)^{n-1}}}e^{-u_nH_{B,(-1)^n}} \Bigg] _{|\lambda=0}\nn\\
&=&\sum_{0\leq i\leq j \leq n} (-1)^{i+j} M_{i,j},
\eea
with 
\bea 
M_{i,i} = \sum_k \frac{g_k^2}{\omega_k^2}(n_{\omega_k}+1)\left(e^{-u_i\omega_k}-e^{-u_{i+1}\omega_k}\right)\left(e^{-(\beta-u_i)\omega_k}-e^{u_{i+1}\omega_k}\right),
\eea
and
\bea 
M_{i,j} \underset{i<j}{=} \sum_k \frac{g_k^2}{\omega_k^2}\Big[\left(e^{-u_i\omega_k}-e^{-u_{i+1}\omega_k}\right)\left(e^{u_j\omega_k}-e^{u_{j+1}\omega_k}\right)(n_{\omega_k}+1) + \left(e^{u_i\omega_k}-e^{u_{i+1}\omega_k}\right)\left(e^{-u_j\omega_k}-e^{-u_{j+1}\omega_k}\right)n_{\omega_k}\Big],
\eea
with the convention $u_0:=\beta$, and $u_{n+1}=0$.
%
One can verify that for $n=1$ we recover the expression of the previous section. We finally  obtain for $t_{+,n}$,
\bea
t_{+,n} &=&\left(\frac{\Delta}{2}\right)^n e^{\epsilon (u_1-u_2+ ... -(-1)^n u_n)}e^{-\lambda^2\sum_{0\leq i\leq j \leq n} (-1)^{i+j} M_{i,j}}|+\ket\bra (-1)^n|.
\eea
Repeating a similar derivation we obtain for $t_{-,n}$,
\be
t_{-,n} = \left(\frac{\Delta}{2}\right)^n e^{-\epsilon (u_1-u_2+ ... -(-1)^n u_n)}e^{-\lambda^2\sum_{0\leq i\leq j \leq n} (-1)^{i+j} M_{i,j}}|+\ket\bra (-1)^n|,
\ee
(only $\epsilon$ is changed to $-\epsilon$).
\end{widetext}
Although we now have an explicit expression of $t_{l,n}$, the nested integrals over the $u_i$ appearing in \eqref{exprTln} are still challenging in general. Thus, as in the main text, we proceed by looking at the high temperature regime, characterized by $\beta \omega_c \ll 1$, where $\omega_c$ stands for the cut-off frequency of the bath spectral density. Consequently, we can expand all exponential functions appearing in the coefficients $M_{i,j}$ (since all $u_i$ are smaller than $\beta$). We obtain the following simple expressions 
\bea
M_{i,i} &=& \frac{Q}{\beta}(u_i-u_{i+1})(\beta-u_i+u_{i+1})\nn\\
M_{i,j} &\underset{i<j}{=} & -2\frac{Q}{\beta}(u_i-u_{i+1})(u_j - u_{j+1})
\eea
Using the above expressions, one can see easily that for $n=1$, we have $\sum_{0\leq i\leq j \leq 1} (-1)^{i+j} M_{i,j} = 4\frac{Q}{\beta}u_1(\beta -u_1)$, recovering a result from the main text. For $n=2$, one obtains  $\sum_{0\leq i\leq j \leq 2} (-1)^{i+j} M_{i,j} = 4\frac{Q}{\beta}(u_1-u_2)(\beta -u_1+u_2)$. More generally, one can show by iteration that for arbitrary $n$, 
\bea
\!\!\!\!\sum_{0\leq i\leq j \leq n}  (-1)^{i+j}M_{i,j} 
&=& 4\frac{Q}{\beta}(u_1-u_2 +u_3 -... u_n)\nn\\
&& ~~\times (\beta -u_1+u_2-u_3 +... u_n).\nn\\
\eea
Finally, we obtain
\bea
&& \!\!\!\!\!\!\!\!\!\!\!\!\!\!\!T_{\pm,n} = \frac{(-\Delta)^n}{2^n} \int_0^\beta\int_0^{u_1}... \int_0^{u_{n-1}}du_1du_2 ...du_n   \nn\\
&& \!\!\!\! e^{\pm\epsilon(u_1-u_2+... u_n)}e^{-4\lambda^2 Q\beta \frac{u_1-u_2+...u_n}{\beta}\left(1-\frac{u_1-u_2+...u_n}{\beta}\right)}\nn\\
&& \!\!\!\!\times|\pm\ket \bra \pm(-1)^n|\nn\\
&& \!\!\!\!= \frac{(-\beta\Delta)^n}{2^n} \int_0^1\int_0^{u_1}... \int_0^{u_{n-1}}du_1du_2 ...du_n \nn\\
&& \!\!\!\! e^{\beta(\pm\epsilon-4\lambda^2Q)(u_1-u_2+... u_n)}e^{4\lambda^2 Q\beta (u_1-u_2+...u_n)^2}|\pm\ket \bra \pm(-1)^n|.\nn\\
\eea
Additionally, we can show that for $n$ impair, $T_{+,n} = e^{\epsilon\beta}T_{-,n}$, which guarantees the Hermicity of $\rho_S^{ss}$. Note that this relation does not hold for $n$ pair, which does not affect the Hermicity of $\rho_S^{ss}$ since corrections of even order affect only the populations. 

\begin{figure*}
(a)\includegraphics[width=80mm]{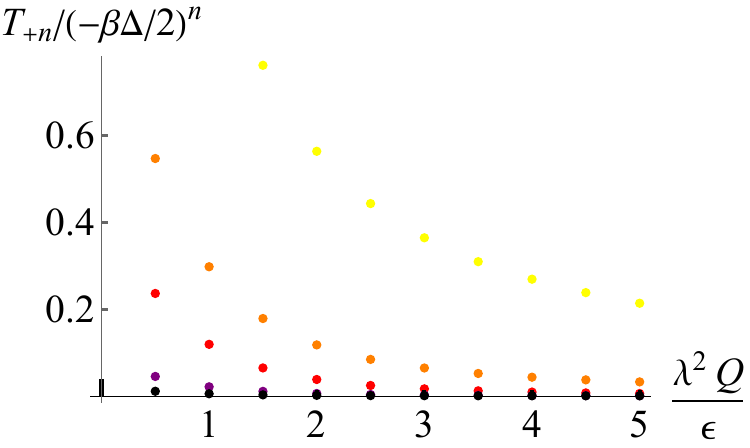} ~~~~~~(b)\includegraphics[width=80mm]{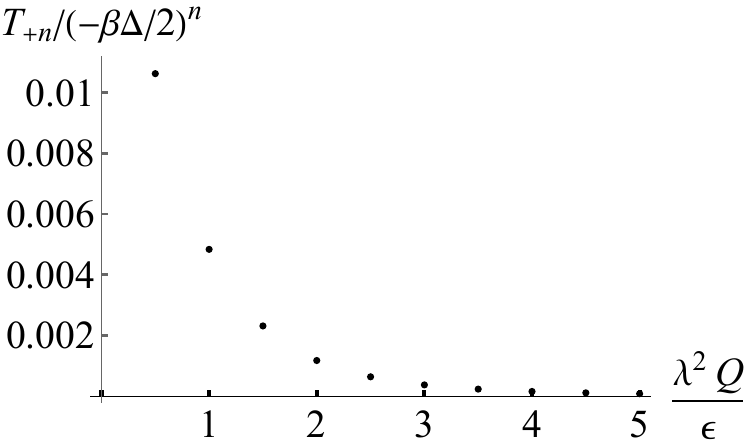}
\caption{ (a) Plots of $T_{+,n}/(-\beta\Delta/2)^n$ in function of $\lambda^2Q$ in unit of $\epsilon$, for $\epsilon\beta=2$, and $n=1$ (yellow dots), $n=2$ (orange dots), $n=3$ (red dots), $n=4$ (purple dots), $n=5$ (black dots). (b) Zoom in of the plot of $T_{+,5}/(-\beta\Delta/2)^5$. }
\label{largelambda}
\end{figure*}

Beyond that, since $(u_1-u_2+... u_n) \in [0;1]$, we have the following simple upper bound,
\bea
|T_{\pm,n}|& \leq& \left(\frac{\beta\Delta}{2}\right)^n \int_0^1\int_0^{u_1}... \int_0^{u_{n-1}}du_1du_2 ...du_n \nn\\
&&\hspace{3.5cm} \times e^{\pm\epsilon\beta(u_1-u_2+... u_n)}\nn\\
\eea
implying 
\bea
|T_{+,n}|& \leq& \frac{1}{n!}\left(\frac{\beta\Delta}{2}\right)^n e^{\epsilon\beta}
\eea
and 
\bea
|T_{-,n}|& \leq& \frac{1}{n!}\left(\frac{\beta\Delta}{2}\right)^n 
\eea
which ensures that the higher order corrections $|T_{\pm,n}|$ vanish quickly as $n$ increases. Additionally, for a fixed $n$, we can show numerically that $|T_{\pm,n}|$ goes to zero as $\lambda$ increases (see plots in Fig.~\ref{largelambda}), confirming that for large~$\lambda$, only the first few orders are enough to obtain a good approximation of $\rho_S^{ss}$.

\subsection{Further analytical comparison between \texorpdfstring{Eq.~\eqref{spin1stapp}}{Eq.(33)} and the result from Timofeev \& Trushechkin}
\label{sec:74}

Here we provide some brief analytical comparison between our results and the one obtained in \cite{Timofeev_2022}.
As detailed in Section~\ref{seccomparison}, the result from \cite{Timofeev_2022} applied to the spin-boson model gives \eqref{rhoMF} in the basis of $\sigma_z$, $\{|+\ket,|-\ket\}$.
In the regime where $\omega_S'\beta \ll1$, we obtain
\bea
\label{rhoMFht}
\rho_{MF} &=& \frac{1}{2}\left( 1 - \frac{\epsilon\beta}{2}\right)|+\ket\bra +|\nn\\
&& +\frac{1}{2}\left( 1 + \frac{\epsilon\beta}{2}\right)|-\ket\bra -| \nn\\
&& -\frac{\Delta\beta}{4}e^{-2\beta\lambda^2Q/3} \Big(|+\ket\bra -| + |-\ket\bra +|\Big).
\eea
For comparison, if we take the limit $\beta\lambda^2Q \ll1$ (and $\epsilon^2\beta/\lambda^2Q \ll 1$) in the expression \eqref{spin1stapp} of $f_{+,-}(\beta)$, we obtain, using $DF(x) = x + {\cal O}(x^3)$ for $x\ll1$,
\bea
\rho_S^{\textrm{ss}} &=& \frac{1}{e^{-\epsilon\beta/2}+e^{\epsilon\beta/2}}\left( e^{-\epsilon\beta/2}|+\ket\bra +| + e^{\epsilon\beta/2}|-\ket\bra -|\right)\nn\\
&&-\frac{\Delta\beta}{4}\left(1+\frac{\epsilon^2\beta}{8\lambda^2Q}\right)\sigma_x.
\eea
Both expressions are equivalent when applying $\beta\lambda^2\beta\ll1 $ to \eqref{rhoMFht}. However, in the opposite limit, when $\beta\lambda^2Q \gg1$, the expression \eqref{spin1stapp} becomes
\bea
\rho_S^{\textrm{ss}} &=& \frac{1}{e^{-\epsilon\beta/2}+e^{\epsilon\beta/2}}\left( e^{-\epsilon\beta/2}|+\ket\bra +| + e^{\epsilon\beta/2}|-\ket\bra -|\right)\nn\\
&&-\frac{\Delta}{8\lambda^2 Q}\sigma_x.
\eea
This is significantly different from \eqref{rhoMF}. While in the above expression the coherences vanish as $1/\lambda^2Q$, they vanish exponentially in \eqref{rhoMF} and \eqref{rhoMFht}.

\subsection{Without initial renormalization}
\label{sec:75}

 As commented above, one can choose to perform the same derivation starting from the ``natural" Hamiltonian 
 \be
 {\cal H}_{SB} = H_S + H_B + \lambda AB,
 \ee
 instead of the renormalized one $H_{SB}$ defined in the main text. Proceeding as previously, we have
\bea
e^{-\beta {\cal H}_{SB}} = e^{-\beta(H_B + \lambda AB)}e^{-{\cal T}\int_0^{\beta} du \tilde {\cal H}_S(u)},
\eea
with 
\be
\tilde {\cal H}_S(u) := e^{u(H_B + \lambda AB)} H_S e^{-u(H_B + \lambda AB)}.
\ee
We now have
\bea
 H_B + \lambda AB  &=& \sum_k \omega_k {\cal D}_k b_k^{\dag}b_k {\cal D}_k^{\dag} -\lambda^2QA^2\nn\\
 &=&\sum_l |a_l\ket\bra a_l| {\cal H}_{B,l} ,
\eea
with ${\cal H}_{B,l} := H_{B,l}  -\lambda^2a_l^2Q =  H_B + \lambda a_l B $, leading to
\bea
e^{u(H_B + \lambda AB)} &=& e^{u \sum_l |a_l\ket\bra a_l| {\cal H}_{B,l}}\nn\\
&=& \sum_l |a_l\ket\bra a_l| e^{u {\cal H}_{B,l}}.
\eea
Then, 
\bea
\tilde H_S(u) &=& \sum_{l,l'} |a_l\ket \bra a_l| e^{uH_{B,l}} H_S |a_{l'}\ket \bra a_{l'}| e^{-uH_{B,l'}}\nn\\
&=& H_S^{\textrm{pop}} + \tilde H_S^{\textrm{coh}}(u),
\eea
where ${\cal H}_S^{\textrm{pop}} =H_S^{\textrm{pop}} = \sum_l h_l |a_l\ket\bra a_l|$ and $\tilde {\cal H}_S^{\textrm{coh}}(u):=  \sum_{l\ne l'} h_{l,l'} |a_l\ket\bra a_{l'}| e^{u(H_{B,l}-\lambda^2a_l^2Q)}e^{-u(H_{B,l'}-\lambda^2a_{l'}^2Q)} =\\ \sum_{l\ne l'} h_{l,l'} |a_l\ket\bra a_{l'}| e^{u{\cal H}_{B,l}}e^{-u{\cal H}_{B,l'}}$.
Again, similarly as previously, we obtain,
\bea
e^{-{\cal T}\int_0^{\beta} du \tilde {\cal H}_S(u)} &=& e^{-\beta H_S^{\textrm{pop}}} e^{-{\cal T}\int_0^\beta du \dtilde {\cal H}_S^{\textrm{coh}}(u)},
\eea
with
\bea
 \dtilde {\cal H}_S^{\textrm{coh}}(u) &:=& e^{u H_S^{\textrm{pop}}} \tilde {\cal H}_S^{\textrm{coh}}(u) e^{-uH_S^{\textrm{pop}}}\nn\\
  &=&  \sum_{l\ne l'} h_{l,l'} e^{u \omega_{l,l'}} |a_l\ket\bra a_{l'}|  e^{u{\cal H}_{B,l}}e^{-u{\cal H}_{B,l'}},
 \eea
arriving at
 \bea
 \rho_{SB}^{\textrm{th}} &=& Z_{SB}^{-1} \sum_l e^{-\beta h_l}|a_l\ket\bra a_l| e^{-\beta {\cal H}_{B,l}}  e^{-{\cal T}\int_0^\beta du \dtilde {\cal H}_S^{\textrm{coh}}(u)}.\nn\\
 \eea
Finally, the main change is that we are led to compute ${\textrm{Tr}}_B\left[e^{-\beta {\cal H}_{B,l}}\right]$ for the zeroth order, and  ${\textrm{Tr}}_B\left[e^{-\beta {\cal H}_{B,l}}  e^{u{\cal H}_{B,l}}e^{-u{\cal H}_{B,l'}}\right]$ for the second order, instead of ${\textrm{Tr}}_B\left[e^{-\beta H_{B,l}}  e^{uH_{B,l}}e^{-uH_{B,l'}}\right]$. We have,
\bea
&&{\textrm{Tr}}_B\left[e^{-\beta {\cal H}_{B,l}}  e^{u{\cal H}_{B,l}}e^{-u{\cal H}_{B,l'}}\right] \nn\\
&=&{\textrm{Tr}}_B\left[e^{-\beta (H_{B,l} - \lambda^2a_l^2Q)}  e^{u(H_{B,l} - \lambda^2a_l^2Q)}e^{-u(H_{B,l'}- \lambda^2a_{l'}^2Q)}\right]\nn\\
&=& {\textrm{Tr}}_B\left[e^{-(\beta-u) H_{B}} e^{-u(H_B + \lambda a_{l',l} B + \lambda^2 a_{l',l}^2 Q)}\right]\nn\\
&&\times e^{\lambda^2Q[\beta a_l^2 + u(a_{l'}^2-a_l^2)]}.
\eea
With that, we obtain
\bea
 \rho_S^{\textrm{ss}} &=&  \sum_l {\mathbf p}_l^{\textrm{ss}} |a_l\ket\bra a_l|  - \sum_{l, l'; l\ne l'}  {\mathbf p}_l^{\textrm{ss}} h_{l,l'} {\mathbf f}_{l,l'}(\beta) |a_{l}\ket\bra a_{l'}|,\nn\\
 \eea 
where ${\mathbf p}_l^{\textrm{ss}}:= e^{-\beta (h_l -a_l^2\lambda^2Q)}/{\cal Z}_S^{\textrm{ss}} = e^{-\beta {\mathbf h}_l }/{\cal Z}_S^{\textrm{ss}}$ is the renormalized population, with ${\mathbf h}_l := h_l -a_l^2\lambda^2Q$ the renormalized ``pseudo-energies" (diagonal elements of $H_S$ in the eigenbasis of $A$), ${\cal Z}^{\textrm{ss}} := \sum_l e^{-\beta {\mathbf h}_l}$, and 
\bea
&&{\mathbf f}_{l,l'}(\beta) := \int_0^\beta du\nn\\
&&\times e^{u\bar \omega_{l,l'}} e^{-\lambda^2 a_{l',l}^2 \int_0^\infty d\omega \frac{J(\omega)}{\omega^2} \frac{e^{\omega\beta/2}+e^{-\omega\beta/2}-e^{\omega (u-\beta/2)}-e^{-\omega(u-\beta/2)}}{e^{\omega\beta/2}-e^{-\omega \beta/2}}},\nn\\
\eea
with $\bar \omega_{l,l'} := {\mathbf h}_l - {\mathbf h}_{l'}$. Additionally, one can also verify the identity ${\mathbf p}_l^{\textrm{ss}} {\mathbf f}_{l,l'}(\beta) = {\mathbf p}_{l'}^{\textrm{ss}}{\mathbf f}_{l',l}(\beta)$.\\

\noindent \emph{Conclusion.} If we do not renormalize the Hamiltonian initially, the expressions are un-changed up to the substitution of $h_l$ by ${\mathbf h}_l$. The renormalization has to happen, either initially, either finally. Note however that $h_{l,l'}$ is not changed. One must be aware of these differences of choice especially when defining the strong coupling regime.

\subsubsection{Approximation of \texorpdfstring{${\mathbf f}_{l,l'}(\beta)$}{fl,l'(beta)}}

Similarly as for ${f}_{l,l'}(\beta)$, when the bath spectral density is such that $J(\omega)$ vanishes for $\omega \geq \omega_c$, where $\omega_c \leq \beta^{-1}$, we have
\bea
{\mathbf f}_{l,l'}(\beta) &\simeq&\int_0^\beta du e^{u \bar\omega_{l,l'}}   e^{-\lambda^2a_{l',l}^2u\left(1-\frac{u}{\beta}\right)Q}\nn\\
&&\hspace{-1cm}=\frac{1}{\lambda |a_{l',l}|}\sqrt{\frac{\beta}{Q}}\Bigg\{{\textrm{DF}}\left[\frac{1}{2\lambda |a_{l',l}|}\sqrt{\frac{\beta}{Q}}(\lambda^2a_{l',l}^2Q-\bar\omega_{l,l'})\right] \nn\\
&&+ e^{\beta\bar\omega_{l,l'}}{\textrm{DF}}\left[\frac{1}{2\lambda|a_{l',l}|}\sqrt{\frac{\beta}{Q}}(\lambda^2a_{l',l}^2Q+\bar\omega_{l,l'})\right]\Bigg\}.\nn\\
\eea
Note that $\lambda^2a_{l',l}^2Q-\bar\omega_{l,l'} = 2\lambda^2Q a_l(a_l-a_{l'}) - \omega_{l,l'} $ and $\lambda^2a_{l',l}^2Q +\bar\omega_{l,l'} = 2\lambda^2Q a_{l'}(a_{l'}-a_{l}) +\omega_{l,l'} $. Thus, in the strong coupling regime when $\lambda^2 Q \gg {\textrm{max}}_{l}~ |h_l|$ and $\lambda^2Q\beta\gg1$, it is still legitimate to approximate the function $DF(x)$ by $1/2x$, which gives,
 \bea\label{ttfsimplified}
{\mathbf f}_{l,l'}(\beta) &=&\frac{1}{2\lambda^2Qa_{l,l'}}\left(\frac{1}{a_l}-\frac{e^{\bar\omega_{l,l'}\beta}}{a_{l'}}\right) \nn\\
&&+\frac{\omega_{l,l'}}{4\lambda^4Q^2a_{l,l'}^2}\left(\frac{1}{a_l^2}-\frac{e^{\bar\omega_{l,l'}\beta}}{a_{l'}^2}\right)+ {\cal O}[(\lambda^2Q/\omega_{l,l'})^{-3}].\nn\\
\eea

\subsection{Identity of first orders}
\label{sec:76}

Here, we compare our general first order result, Eqs.~\eqref{final2dorder} and \eqref{genf}, with the general first order expression for the steady state coherences obtained in Eq.~(56) of \cite{Trushechkin_2021}. This expression was obtained through projection operator techniques \cite{FrancescoBook,Nakajima_1958,Zwanzig_1960}, by choosing the operator ${\cal P}$ projecting the system's state onto the diagonal subspace (in the eigenbasis of the unperturbed Hamiltonian). Thus, the non-diagonal elements can be obtained by establishing and solving the dynamics associated with the complementary operator $1-{\cal P}$ (section IV.A. of \cite{Trushechkin_2021}). The steady state coherences are then deduced by taking the time to infinity. One should keep in mind that the master equation derived in the strong-decoherence limit in \cite{Trushechkin_2021} is valid when the bath spectral density is such that $\lim_{\omega \rightarrow +\infty} \frac{J(\omega)}{\omega}$ is non-zero (and possibly infinite). This condition, satisfied by usual spectral densities, guarantees that the unperturbed dynamics in \cite{Trushechkin_2021} leads to full decoherence, which is the starting point of the projection technique used therein. 

To make the comparison easier, we re-write in the following the results of \cite{Trushechkin_2021} using the notations we have been using here. Additionally, in order to simplify the presentation, we consider a simple coupling of the form $V_I = \lambda AB$ instead of the general one $V_I = \sum_\alpha A_\alpha B_\alpha$ considered in \cite{Trushechkin_2021}. With that, in the eigenbasis $\{|a_l\ket\}_l$ of $A$, the result of \cite{Trushechkin_2021} takes the form
\bea\label{exprcohME}
\rho_{l,l'}^{\textrm{ss,ME}} &:=& \bra a_l|\rho_S^{\textrm{ss,ME}} |a_{l'}\ket\nn\\
 &=& h_{l,l'}\Big[i  {\mathbf p}_l^{\textrm{ss}} \int_0^{\infty} d\tau e^{-\lambda^2 a_{l',l}^2[G^*(\tau) -i\tau Q]}e^{-i\bar\omega_{l,l'} \tau} \nn\\
 &&  - i {\mathbf p}_{l'}^{\textrm{ss}}  \int_0^{\infty} d\tau e^{-\lambda^2 a_{l,l'}^2[G(\tau) +i\tau Q]}e^{-i\bar\omega_{l,l'} \tau} \Big]\nn\\
\eea
with ${\mathbf p}_l^{\textrm{ss}}:= e^{-\beta (h_l -a_l^2\lambda^2Q)}/{\cal Z}_S^{\textrm{ss}} = e^{-\beta {\mathbf h}_l }/{\cal Z}_S^{\textrm{ss}}$ is the renormalized population introduced in Section~\ref{sec:75}, with ${\mathbf h}_l := h_l -a_l^2\lambda^2Q$ the renormalized ``pseudo-energies", and ${\cal Z}^{\textrm{ss}} := \sum_l e^{-\beta {\mathbf h}_l}$. Additionally, we defined $G(\tau) := \int_0^\tau d s_1\int_0^{s_1}ds_2 c_B(s_2)$, and 
\bea
c_B(s) =  {\textrm{Tr}}_B [e^{i H_B s} B e^{-i H_B s} B\rho_B^{\textrm{th}}]. 
\eea
The superscript ``ME" refers to the master equation nature of the derivation \cite{Trushechkin_2021}.

Our result can be expressed as (using the results of the derivation with no initial renormalization of the Hamiltonian, detailed in Section~\ref{sec:75}, since it is the choice made in \cite{Trushechkin_2021}),
 \bea
\!\!\!\!\!\!\!\!\rho_{l,l'}^{\textrm{ss}}   &:=& \! \bra a_l|\rho_S^{\textrm{ss}} |a_{l'}\ket \nn\\
&=& \!\! - h_{l,l'}{\mathbf p}_l^{\textrm{ss}} \int_0^{\beta} du e^{-\lambda^2a_{l,l'}^2[G(-iu) + uQ]}e^{u\bar\omega_{l,l'}}\nn\\
&=& \!\! -h_{l,l'}{\mathbf p}_{l'}^{\textrm{ss}} \int_0^{\beta} du e^{-\lambda^2a_{l',l}^2[G(-iu) + uQ]}e^{u\bar\omega_{l',l}},
\eea
where we used in the second line the identity  ${\mathbf p}_l^{\textrm{ss}} {\mathbf f}_{l,l'}(\beta) = {\mathbf p}_{l'}^{\textrm{ss}}{\mathbf f}_{l',l}(\beta)$, shown in Section~\ref{sec:72}.
Now, let us consider the integral of the function $e^{-\lambda^2a_{l',l}^2[G(-iu) + uQ]}e^{u\bar\omega_{l',l}}$ extended to the complex plan along the contour ${\cal C}$ defined as $u: 0\rightarrow \beta \rightarrow \beta + i x \rightarrow ix \rightarrow 0$, where $x$ is a positive number that later will be taken to infinity. Since it is a closed contour, the integral is equal to zero (as long as $G(u)$ is an analytic function).
 Additionally, when the bath spectral density satisfies $\lim_{\omega \rightarrow +\infty} \frac{J(\omega)}{\omega^2}>0$, the real part of the function $G(x)$ tends to $+\infty$ as $x$ increases. The condition $\lim_{\omega \rightarrow +\infty} \frac{J(\omega)}{\omega^2}>0$ is precisely the condition of validity of the results in \cite{Trushechkin_2021}. This is not a coincidence. The validity of the master equation in \cite{Trushechkin_2021} relies on the full decoherence of the unperturbed dynamics, which is guaranteed as long as $G(x)$ tends to $+\infty$ as $x$ increases, which itself results in the condition $\lim_{\omega \rightarrow +\infty} \frac{J(\omega)}{\omega^2}>0$ on the bath spectral density.
Then, the integral of $e^{-\lambda^2a_{l',l}^2[G(-iu) + uQ]}e^{u\bar\omega_{l',l}}$ tends to zero on the segment $\beta + ix \rightarrow i x$ for $x \rightarrow +\infty$.
%
%
 As a result, we have
\bea
\rho_{l,l'}^{\textrm{ss}}   &=&  h_{l,l'}{\mathbf p}_{l'}^{\textrm{ss}} \lim_{x\rightarrow +\infty}\int_\beta^{\beta +ix} du e^{-\lambda^2a_{l',l}^2[G(-iu) + uQ]}e^{u\bar\omega_{l',l}}\nn\\
&& +  h_{l,l'}{\mathbf p}_{l'}^{\textrm{ss}} \lim_{x\rightarrow +\infty}\int_{+ix}^{0} du e^{-\lambda^2a_{l',l}^2[G(-iu) + uQ]}e^{u\bar\omega_{l',l}}\nn\\
&=&  i h_{l,l'}{\mathbf p}_{l'}^{\textrm{ss}} \int_0^{+\infty} d\tau e^{-\lambda^2a_{l',l}^2[G(\tau-i\beta) + (i\tau+\beta)Q]}e^{(i\tau+\beta)\bar\omega_{l',l}}\nn\\
&& - i h_{l,l'}{\mathbf p}_{l'}^{\textrm{ss}} \int_{0}^{+\infty} d\tau e^{-\lambda^2a_{l',l}^2[G(\tau) +i\tau Q]}e^{i\tau\bar\omega_{l',l}}\nn\\
&=&  i h_{l,l'}{\mathbf p}_{l}^{\textrm{ss}} \int_0^{+\infty} d\tau e^{-\lambda^2a_{l',l}^2[G^{*}(\tau) -i\tau Q]}e^{-i\tau\bar\omega_{l,l'}}\nn\\
&& - i h_{l,l'}{\mathbf p}_{l'}^{\textrm{ss}} \int_{0}^{+\infty} d\tau e^{-\lambda^2a_{l',l}^2[G(\tau) +i\tau Q]}e^{-i\tau\bar\omega_{l,l'}}\nn\\
\eea
which is precisely equal to the expression~\eqref{exprcohME} of $\rho_{l,l'}^{\textrm{ss,ME}}$. Note that in the last line we used the identities ${\mathbf p}_{l'}^{\textrm{ss}}e^{\beta\bar\omega_{l',l}} ={\mathbf p}_{l}^{\textrm{ss}}$ and $G(\tau-i\beta) + (i\tau+\beta)Q = G^{*}(\tau)-i\tau Q$ (as well as $\omega_{l',l}=-\omega_{l,l'}$). This concludes the proof that the expression derived in this paper coincides with the expression obtained in \cite{Trushechkin_2021}, although they have been obtained from very different methods and starting point. The equivalence of the expressions is guaranteed as soon as the results in \cite{Trushechkin_2021} are valid (namely, as soon as $\lim_{\omega \rightarrow +\infty} \frac{J(\omega)}{\omega^2}>0$).
As reminded in the main text, this is an important step forward since it proves that the mean force Gibbs state is indeed the actual steady state (at least up the first order) even when the system interacts \emph{strongly} with the thermal bath.

\section*{Acknowledgments}

I would like to thank Anton Trushechkin for very interesting and helpful discussions on this topic. I am also grateful to the Editors for their constructive comments, participating in improving the quality of the paper.

\section*{Funding}

This work was supported by funding from the National Institute for Theoretical Physics (NITheP) of the Republic of South Africa.\\

\bibliography{references}
\bibliographystyle{quanta}

\end{document}